\documentclass[12pt,letterpaper,english,aps,preprint,nofootinbib]{revtex4}
\usepackage[T1]{fontenc}
\usepackage[latin9]{inputenc}
\usepackage{verbatim}
\usepackage[dvips]{graphicx}

\makeatletter

\makeatletter

\usepackage{babel}
\makeatother

\begin{document}
\newcommand{\oalphas}{O(\alpha_{s})}

\newcommand{\met}{\not\!\! E_{T}}

\title{Colliding neutrino beams}

\author{Reinhard Schwienhorst}

\email{schwier@pa.msu.edu}

\affiliation{Department of Physics $\&$ Astronomy, Michigan State University,
East Lansing, MI 48824, USA}

\date{\today}

\begin{abstract}
From several neutrino oscillation experiments, we understand now that
neutrinos have mass. However, we really don't know what mechanism
is responsible for producing this neutrino mass. Current or planned
neutrino experiments utilize neutrino beams and long-baseline detectors
to explore flavor mixing but do not address the question of the origin
of neutrino mass. In order to answer that question, neutrino interactions
need to be explored at much higher energies. This paper outlines a
program to explore neutrinos and their interactions with various particles
through a series of experiments involving colliding neutrino beams.
\end{abstract}
\maketitle

\section{Introduction\label{sec:Introduction}}

Neutrinos were first introduced to preserve the law of energy conservation
in nuclear beta decay. The first experimental studies of neutrinos
came with the advent of neutrino beams~\cite{Danby:1962nd}. The
understanding of neutrinos had been limited to their role in weak
interactions, where they participate as massless left-handed leptons.
In the Standard Model (SM), neutrinos have been massless for the longest
time. Several experiments involving atmospheric~\cite{Fukuda:1998mi},
solar~\cite{Davis:1968cp,SNO:2002}, and accelerator neutrinos~\cite{Aliu:2004sq,Michael:2006rx}
have changed that picture recently. We now understand that neutrinos
have masses, albeit small, and that the mass Eigenstates are not Eigenstates
of the weak interaction and thus neutrinos undergo flavor oscillations.
This picture is rather similar to the quark sector where mixing between
generations has been studied for several decades. In the quark sector,
experiments are now starting to explore not only details about this
flavor mixing but also the actual mechanism that produces quark masses,
i.e. the Higgs mechanism. Within the SM, the Higgs mechanism is responsible
for electroweak symmetry breaking and fermion masses~\cite{Higgs:1964ia}.
But the Higgs boson hasn't been observed yet and is not the only possible
mechanism leading to particle masses. There are many other theories
that could explain electroweak symmetry breaking and the generation
of particle masses. Searches are currently underway at the Fermilab
Tevatron for direct production of the Higgs boson, both within the
SM~\cite{Abazov:2004jy,Abazov:2005un,Abazov:2006hn,Abazov:2006pt,Abulencia:2006aj,Acosta:2005ga},
and from theories beyond the SM~\cite{Abazov:2004au,Abazov:2005yr,Abazov:2006ih,Abulencia:2005jd,Abulencia:2005kq,Acosta:2004uj,Acosta:2005bk,Acosta:2005np}.
Moreover, the LHC proton-proton collider at Cern will start operating
soon and is expected to reveal the electroweak symmetry breaking mechanism~\cite{unknown:1994pu,unknown:1999fq,unknown:1999fr,Armstrong:1994it}.

For neutrinos, no such program to understand the origin of their mass
exists or has even been thought of. There is no theoretical physics
reason for the same mass generation mechanism to apply to neutrinos
and other fermions of the SM~\cite{Mohapatra:2006gs,Caldwell:1993kn,Ma:1995gf,Berezhiani:1995yi,Benakli:1997iu}.
It is not clear at all that neutrinos acquire mass in the same way
as quarks because the mass scales are so different between neutrinos
and quarks and even between neutrinos and their partner leptons. And
not only are neutrino masses significantly smaller than their partner
lepton masses, but only left-handed neutrinos participate in weak
charged current interactions~\cite{PDBook}. 

The current focus for experimental neutrino physics is to understand
the parameters of neutrino flavor oscillations~\cite{Barger:2003qi}.
Several experiments are currently running or approved that will study
neutrino oscillations in more detail: K2K~\cite{Aliu:2004sq}, Minos~\cite{Ables:1995wq},
and Cern to Gran~Sasso~\cite{Acquistapace:1998rv}. All these experiments
intend to address the lepton-equivalent of the CKM matrix. However,
none of them can reveal the actual mechanism that is responsible for
neutrino masses. Some ideas of looking for production of Majorana
neutrinos at hadron colliders such as the LHC have been proposed~\cite{Han:2006ip},
but those look for new particles produced in the weak interaction
rather than actually testing the neutrino mass generation mechanism.
Similarly, there are several experiments looking for neutrinoless
double-beta decay, which would provide a measurement of the absolute
scale of neutrino masses and establish the existence of Majorana neutrinos~(see
\cite{Avignone:2007fu} and references therein). However, all of
them look for evidence of a Majorana neutrino participating in the
weak interaction and none of them probe the actual neutrino mass generation
mechanism directly.

Observing the mass generation mechanism directly means looking for
new particles that couple to neutrinos and possibly observing Yukawa
interactions of neutrinos. And so far, there has not been any idea
or proposal to accomplish this. In fact, the only neutrino interactions
that have been observed so far have been weak interactions, neutrino
couplings to the $W$~and $Z$~bosons. Of course, within the SM
neutrinos are massless and no other interactions exist. But since
neutrinos have non-zero mass, neutrino interactions other than through
the weak interaction must exist. They arise from physics beyond the
SM, involving new particles, symmetries, and interactions. Moreover,
since they have mass, neutrinos may have V+A couplings besides the
SM V-A coupling. In order to reach the energies required to observe
such interactions and produce most of these new particles, high-energy
neutrino beams are required, and neutrino collisions with sufficient
center of mass energy to produce these particles directly. In this
paper, we describe how colliding neutrino beams can be used to explore
neutrino interactions at the highest energies and possibly reveal
the neutrino mass generation mechanism. 

Neutrino beams were initially proposed to study the weak interaction~\cite{Schwartz:1960hg,Pontecorvo:1959sn,Danby:1962nd}.
Neutrino beams have also been used to study the structure of nuclear
matter. High-statistics neutrino experiments have mapped out the parton
composition of nuclei~\cite{Onengut:2005kv,Seligman:1997mc,Aivazis:1993pi,Rabinowitz:1993xx,Abramowicz:1985xg,Jones:1987gk}.
These experiments have always used complex nuclei in order to have
large target mass. Similarly, neutrino-electron scattering results
have been obtained by directing neutrino beams onto targets and then
extracting electron scattering events from the large background of
nuclear scattering events~\cite{Vogel:1989iv}.

The configurations outlined below collide neutrinos with protons,
electrons, muons and other particles, many for the first time. We
will show that while neutrino energies that can be reached with current
accelerators are sufficient to study neutrino interactions at the
highest energies, currently achievable luminosities are not. Future
accelerators such as a neutrino collider based on muons storage rings
are required to achieve significantly higher luminosities and study
neutrino-neutrino interactions for the first time.

\section{Physics with colliding neutrinos}

The experimental focus for neutrinos so far has been on relatively
low energies compared to beams of other particles (electrons or protons).
A neutrino energy range of a few GeV is well suited to study neutrino
oscillations, but doesn't help much when trying to observe neutrino
collisions or reveal the underlying mechanism that generates neutrino
masses. In order to explore potential new interactions and particles,
high energy neutrino beams are required so that the center-of-mass
energy is sufficient to produce possible new particles directly. Within
the SM, the largest interaction rate in neutrino colliders is due
to $W$~boson exchange, between neutrinos from one beam and a particle
(proton, electron, muon, or neutrino) from the other beam. The cross
section for these electroweak charged current interactions is small,
thus primary beams of unprecedented intensity will be needed. The
interactions are straightforward to detect in a typical high-energy
detector because each interaction produces a high-momentum muon (or
electron) that tags the interaction.

Colliding neutrinos with beams of other particles has two main advantages
over fixed target experiments. The first one is the obvious increase
in the center-of-mass energy. For a neutrino beam incident on a fixed
target, the center-of-mass energy is roughly the square root of the
beam energy. By contrast, colliding beams have a center-of-mass energy
of twice the beam energy. While low energy beams and collisions can
be used to study properties of matter and investigate neutrino mixing,
they are not sufficient to produce heavy objects like $W$~or $Z$~bosons
directly. In order to reach the energies required to produce not only
the heavy bosons but also potentially new (and heavier) particles,
much higher center-of-mass energies are required. The second advantage
is that the interacting particles are only those present in the beams
rather than the complex nuclear structure of a fixed target. In particular,
colliding neutrino and electron beams allows for the first low-background
study of neutrino-electron interactions without having to worry about
nuclear matter. And finally, colliding a neutrino beam with a muon
beam allows for the first ever study of neutrino-muon collisions,
something that is not possible in a fixed-target experiment. 

These collisions of neutrinos with other particles can be useful to
study not just the weak interaction or properties of the particles
involved but also to search for new physics in the coupling between
neutrinos and other matter.

\subsection{Neutrino-quark interactions}

In neutrino-proton interactions, it is actually the $W$~boson that
probes the structure of the proton, equivalent to the HERA electron-proton
collider where a photon probes the structure of the proton~\cite{Abramowicz:1998ii}.
Along these lines, a neutrino beam colliding with a proton beam can
serve to measure parton distribution functions of the proton. Moreover,
one is not restricted to proton beams and could also collide neutrinos
with pion or kaon beams, for example. Such a collider enables the
first direct study of the structure of mesons from a weak interaction
perspective. And one could in principle also imagine colliding neutrino
beams with heavy ion beams, although heavy nuclei have already been
studied in detail with neutrino beams incident on nuclear targets.

\begin{figure}
\includegraphics[scale=0.6]{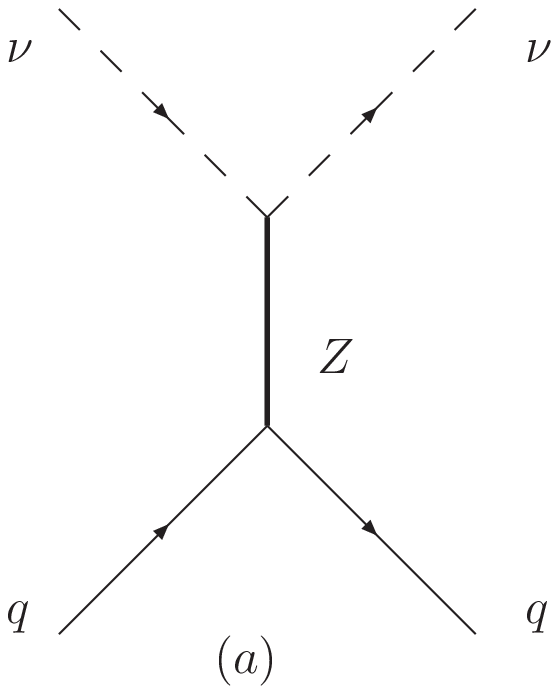}
\includegraphics[scale=0.6]{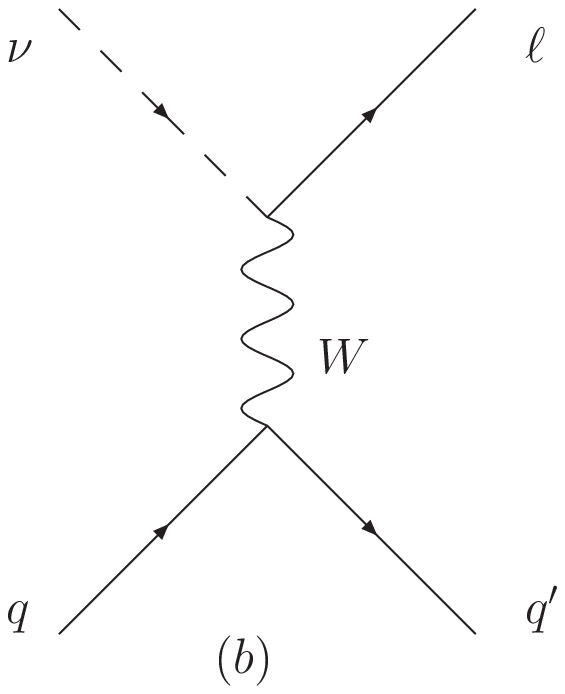}

\caption{Feynman diagrams for neutrino-quark collisions in the SM. Shown are
(a) the $t$-channel exchange of a $Z$~boson and (b) the $t$-channel
exchange of a $W$~boson.\label{fig:FeynmanNeuProt}}

\end{figure}

Fig.~\ref{fig:FeynmanNeuProt} shows the relevant Feynman diagrams
for SM neutrino-proton collisions. A quark from the proton exchanges
a $Z$~or $W$~boson with the neutrino, leading to a final state
of only a quark jet or a lepton plus quark jet. Experimentally, the
lepton produced in the $W$~boson exchange can be used to identify
the interaction, making it straightforward to select neutrino-quark
charged current scattering events.

\subsection{Neutrino-lepton interactions}

\begin{figure}
\includegraphics[scale=0.6]{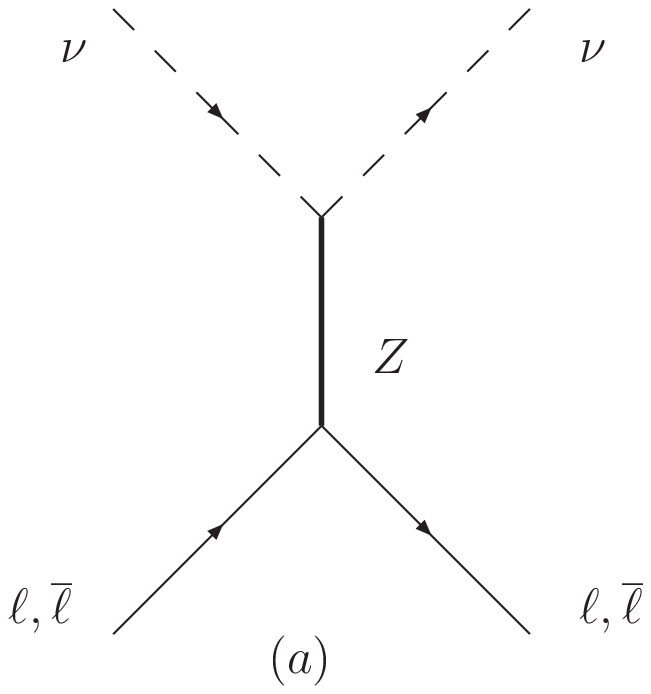}
\includegraphics[scale=0.6]{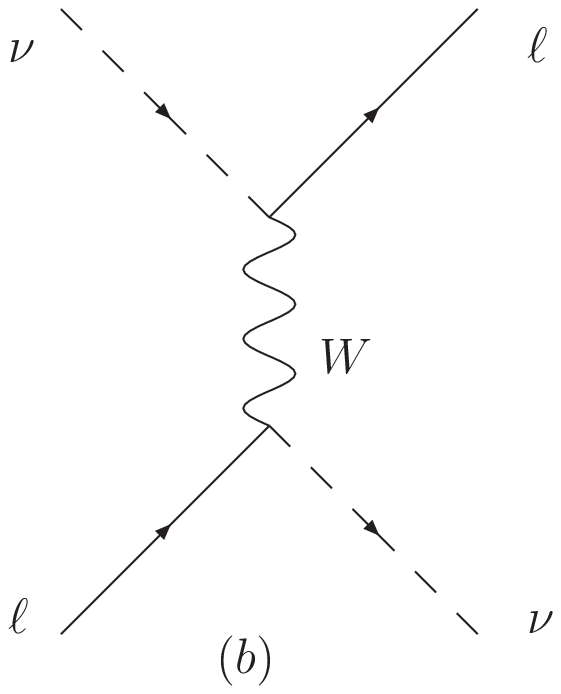}
\includegraphics[scale=0.6]{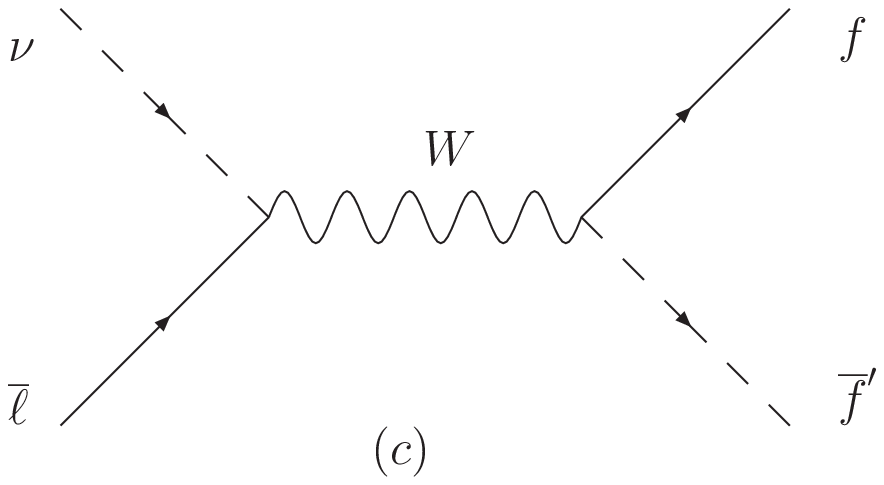}

\caption{Feynman diagrams for neutrino-lepton collisions in the SM. Shown are
(a) the $t$-channel exchange of a $Z$~boson, (b) the $t$-channel
exchange of a $W$~boson, and (c) the $s$-channel production of
a $W$~boson with decay to a fermion-antifermion' pair.\label{fig:FeynmanNeuLep}}

\end{figure}

Within the SM, neutrinos interact with leptons either through charged
current $W$~boson exchange or through neutral current $Z$~boson
exchange. Several different Feynman diagrams contribute, as shown
in Fig.~\ref{fig:FeynmanNeuLep}. 

The exchange of a $Z$~boson is always possible for any combination
of beam particles (neutrino or antineutrino colliding with electron
or positron, Fig.~\ref{fig:FeynmanNeuLep}~(a)). If the neutrino
beam and the lepton beam are both particles (or both antiparticles),
then the $t$-channel exchange of a $W$~boson is also possible (Fig.~\ref{fig:FeynmanNeuLep}~(b)).

If the neutrino beam and the lepton beam are particle-antiparticle
pairs of the same flavor, then $s$-channel annihilation into a $W$~boson
is allowed, which leads to a final state typical for $W$~boson decay
of either lepton-neutrino pair or quark-jet pair (Fig.~\ref{fig:FeynmanNeuLep}~(c)).
Since neutrinos undergo flavor mixing, this annihilation also occurs
for neutrino and lepton beams of different families.

For example, in a muon neutrino-antimuon collider, the muon neutrino
annihilates with the antimuon to form a $W$~boson. The decay of
this $W$~boson produces the typical signature of either two quarks
or one lepton and missing transverse energy, both of which are straightforward
to identify in a detector. In both cases the $W$~boson can be reconstructed
from the final state particles, although in the case of the lepton+neutrino
final state only the transverse components of the neutrino momentum
can be reconstructed due to the wide spread of beam neutrino momenta.
There are also $t$-channel exchanges of $Z$~bosons and $W$~bosons
between the neutrino and the lepton (Figs.~\ref{fig:FeynmanNeuLep}~(a)
and~(b)). These interactions also produce a final state of a lepton
and neutrino, but both lepton and neutrino are more in the forward
direction and there is no resonance invariant mass peak. Thus these
interactions are harder to identify in a detector.

Models of new physics that have higher order symmetries result in
additional heavy $W'$~and $Z'$~bosons~\cite{Pati:1974yy}. If
these new bosons couple only to the lepton sector and not the quark
sector, then a neutrino-lepton collider is especially sensitive to
them. The interactions are the same as shown in Fig.~\ref{fig:FeynmanNeuLep},
but replacing the exchange $W$~and $Z$~bosons by heavier $W'$~and
$Z'$~bosons. 

\begin{figure}
\includegraphics[scale=0.6]{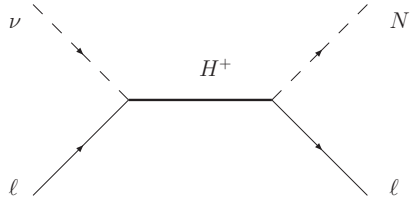}

\caption{Example Feynman diagram for $s$-channel production of a charged Higgs
boson which subsequently decays to a heavy neutrino and a lepton.\label{fig:FeynmanNeuLepNP}}

\end{figure}

A neutrino-lepton collider is able to produce heavy Dirac or Majorana
neutrinos through the exchange of neutral or charged Higgs bosons
that arise in many models of new physics, as shown in Fig.~\ref{fig:FeynmanNeuLepNP}~\cite{Mohapatra:1979ia,Mohapatra:1980yp,PhysRevLett.86.2502}.
The figure shows the production of a charged Higgs boson which then
decays to a heavy neutrino and a lepton. These interaction vertices
are similar to those possible at electron colliders~\cite{Heusch:2000cp,Vuopionpera:1994kr},
except that here a neutrino appears in the initial state. If the right-handed
neutrino has weak interactions, then the production of a heavy right-handed
neutrino in the decay of a $W$~boson is also possible.

\subsection{Neutrino-neutrino interactions}

Colliding neutrinos with other neutrinos is not only a sensitive probe
to new physics involving neutrinos but also allows for SM measurements
that would otherwise not be possible. In particular, a measurement
of the total $Z$~boson production rate in neutrino collisions provides
a direct measurement of the neutrino coupling to the $Z$~boson.
This coupling has only been measured indirectly so far, in single-photon
production~\cite{Abreu:1996vd,Achard:2003tx} as well as in $Z$~boson
decays through a measurement of the total $Z$~decay width and in
neutrino-nucleon neutral current scattering experiments~\cite{Z-Pole}.
A neutrino-antineutrino collider probes this coupling directly through
the annihilation of neutrino and antineutrino into a $Z$~boson. 

\begin{figure}
\includegraphics[scale=0.6]{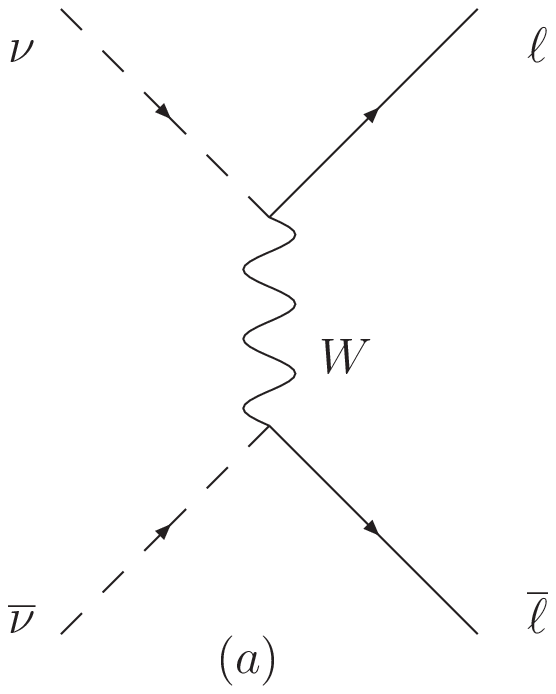}
\includegraphics[scale=0.6]{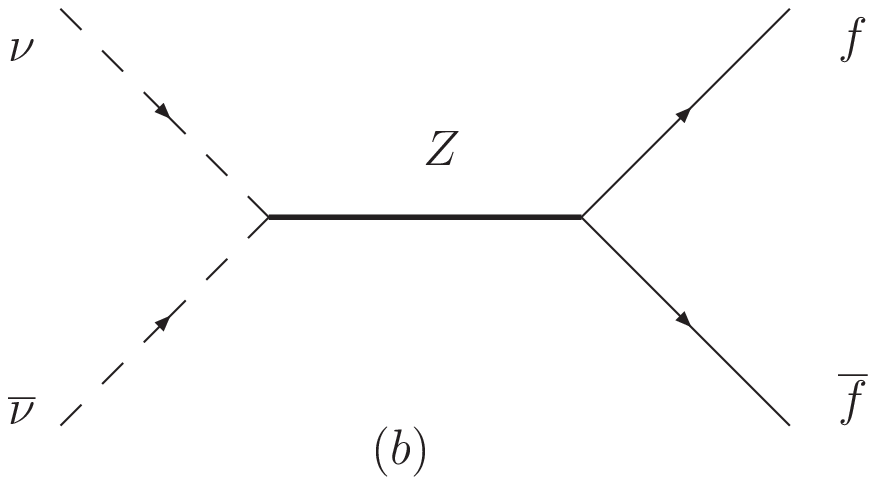}

\caption{Feynman diagrams for neutrino-neutrino collisions in the SM. Shown
are (a) the $t$-channel exchange of a $W$~boson and (b) the $s$-channel
production of a $Z$~boson with subsequent decay to a fermion pair.\label{fig:FeynmanNeuNeu}}

\end{figure}

A large event rate in neutrino-antineutrino collisions is due to $W$~boson
exchange between two neutrinos. This results in a final state of two
oppositely charged leptons which are straightforward to observe. This
is also the final state that can be used to establish the existence
of collisions when neutrino colliders are turned on for the first
time and the luminosity is relatively low. There is also a neutral
current $Z$~boson $t$-channel exchange between the two neutrinos,
but since the final state consists of two neutrinos, this interaction
will not be observable.

\begin{figure}
\includegraphics[scale=0.6]{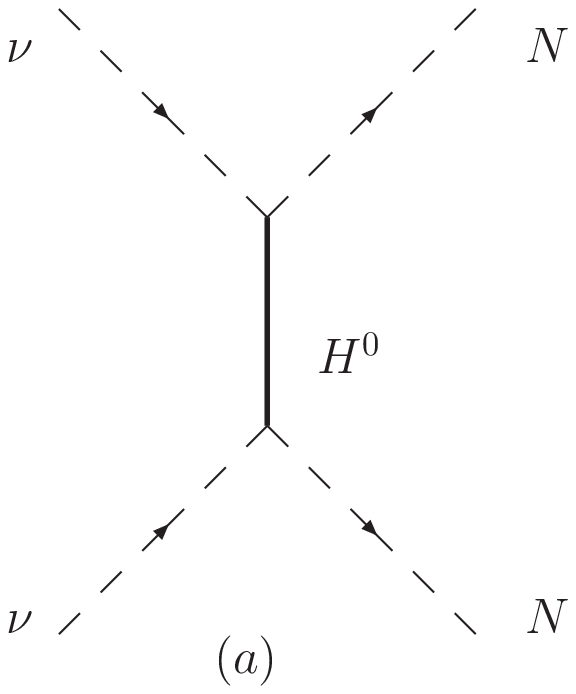}
\includegraphics[scale=0.6]{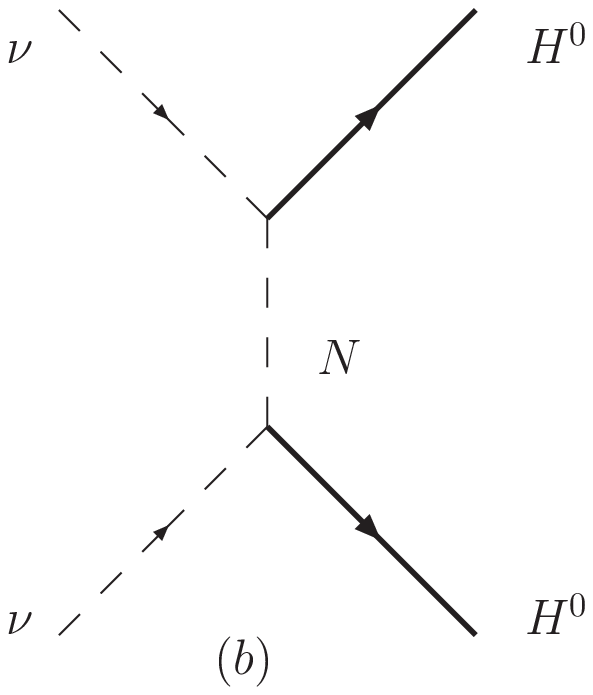}
\includegraphics[scale=0.6]{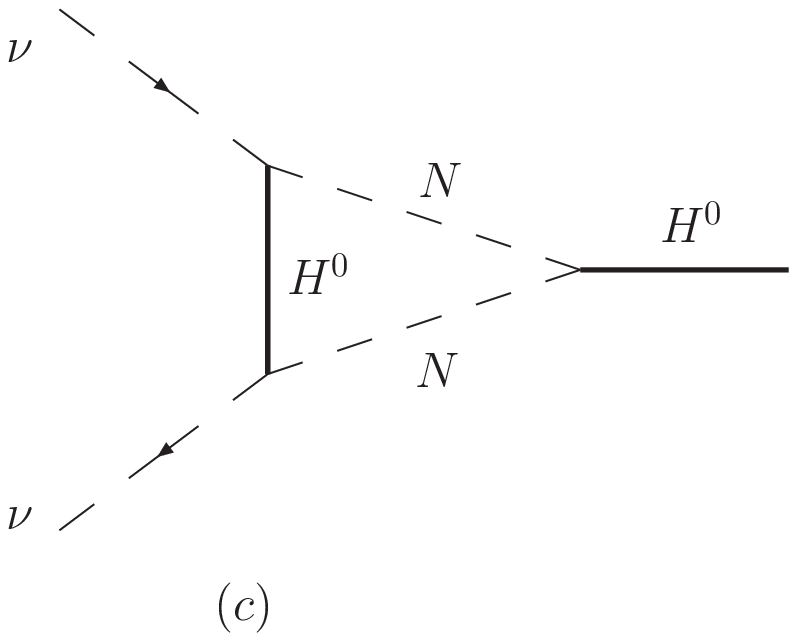}

\caption{Example Feynman diagrams for neutrino-neutrino collisions involving
heavy neutrinos. Shown are (a) the $t$-channel exchange of a Higgs
boson, (b) the $t$-channel exchange of a heavy neutrino, and (c)
the resonance production of a Higgs boson through a loop involving
the Higgs boson and heavy neutrinos.\label{fig:FeynmanNeuNeuNP}}

\end{figure}

All of the interactions and beam configurations discussed so far are
sensitive to new physics in the neutrino sector. But colliding neutrinos
with antineutrinos has more potential to observe new physics due to
the possibility of producing new particles. Just as in the case of
a neutrino-lepton collider, heavy copies of the SM $W$~and $Z$~bosons
can contribute through diagrams as shown in Fig.~\ref{fig:FeynmanNeuNeuNP}.
Two more examples are shown in Fig.~\ref{fig:FeynmanNeuNeuNP}. In
Fig.~\ref{fig:FeynmanNeuNeuNP}~(a), a new Higgs boson that couples
mainly to neutrinos ($H^{0}$) is being exchanged between the two
incoming neutrino beams, leading to a final state of two heavy neutrinos.
Similarly, Fig.~\ref{fig:FeynmanNeuNeuNP}~(b) shows the exchange
of a heavy neutrino leading to a final state of two Higgs bosons.
These two processes are in particular interesting to distinguish whether
the heavy neutrino is of Dirac type or Majorana type. If it is a Majorana
neutrino, then it couples both to neutrinos and antineutrinos, and
thus the final state heavy neutrinos in Fig.~\ref{fig:FeynmanNeuNeuNP}~(a)
are indistinguishable, leading to lepton-flavor violation and final
states of like-sign events. Along the same lines, the incoming neutrinos
don't have to be of opposite flavor, thus the cross section is enhanced
at a neutrino-neutrino collider.

Exploring the mechanism that gives neutrinos masses should start after
the LHC will hopefully have revealed the origin of quark masses. A
new model of quark masses should emerge from the LHC measurements
and possibly from precision measurements at the next linear collider.
Any neutrino collider such as described in this paper will not be
set up until after the LHC has yielded results. Thus, an important
aspect of neutrino colliders will be to observe interactions predicted
by the new theory and to check if neutrinos fit into the picture that
will emerge from the LHC era. We don't know yet what the LHC will
bring, but new physics is generally expected to appear at the TeV
scale, within reach of the LHC, and quite possibly within reach of
a neutrino collider. And producing heavy neutrinos, Higgs bosons,
and other new particles in neutrino colliders is really the only
method to observe the neutrino mass generation mechanism directly.
There may also be surprises awaiting at higher energies where the
SM is expected to break down. 

Moreover, neutrino beams are 100\% polarized by nature, and that feature
can be exploited. Angular correlations between the two leptons produced
in the $W$~boson exchange process can for example be used to study
the nature of the weak interaction and the coupling of the $W$~boson
to neutrinos. They can also be used to look for right-handed neutrinos
in the beam.

\section{\label{sec:Neutrino-Beams}Neutrino beams}

Neutrinos by themselves cannot be accelerated, accumulated, or stored
in a magnetic storage ring. In order to produce a neutrino beam, it
is necessary to accumulate, store, and accelerate charged particles
which then in turn produce the neutrino beam. Once produced, the neutrino
beam cannot be manipulated further.

\subsection{Neutrinos from proton beams}

All of the neutrino beams used in modern oscillation experiments are
produced in the same fashion: High-energy protons hit a stationary
target and the resulting pions and kaons are allowed to decay to muons
and muon neutrinos in a long decay pipe. Some of the muons decay further
to electrons, electron neutrinos, and muon antineutrinos. Strong magnetic
fields are be used to focus the pions and select pions of a certain
charge and momentum. The dominant decay process is 
$\pi^{+}\rightarrow\mu^{+}\bar{\nu}_{\mu}$.
Some of these muons decay further, $\mu^{+}\rightarrow e^{+}\nu_{\mu}\bar{\nu}_{e}$,
which means a neutrino beam produced by protons always contains a
mixture of neutrinos. Nevertheless, selecting positively charged pions
results in a beam consisting predominantly of anti-muon-neutrinos,
and selecting negatively charged pions results in a beam consisting
predominantly of muon-neutrinos. The neutrino beam energy spectrum
is not uniform. The spectrum is rather broad and varies with neutrino
flavor due to the decay chains involved. The mean neutrino energy
is much lower than the incident proton beam energy. 

For wide-band neutrino beams used in neutrino-nucleon scattering
experiments, neutrino interactions have been observed for neutrino
energies up to about half the incident proton energy~\cite{McFarland:1995sr}.
Despite the relatively small number of these high energy neutrinos,
their interactions can nevertheless been observed because the linear
rise in cross section with neutrino energy compensates for the falling
spectrum. The reach in neutrino energy and thus the physics potential
for neutrino beams is driven by the incident proton energy. Still,
most of the interacting neutrinos have less than about 10\% of the
initial proton energy.

Since this article focuses on colliding high-energy beams, only the
neutrino beams at Fermilab and at Cern will be addressed. While there
are more neutrino beams in operation or planned, at KEK~\cite{Noumi:1997}
and other facilities, those all utilize lower energy to explore specific
aspects of neutrino flavor oscillations.

At Fermilab, the current Minos neutrino beam is produced from 120~GeV
protons extracted from the main injector incident on a graphite target.
The neutrino beam energy can be selected with two focusing horns.
In what is called the high-energy running mode, this results in typical
mean neutrino energies of around 10~GeV, with neutrino energies extending
up to about 20~GeV~\cite{Crane:1995ky}. 

At Cern, it is planned for the CNGS neutrino beam to be produced by
400~GeV protons extracted from the SPS incident on a graphite target.
Subsequent focusing results in a mean neutrino energy of about 25~GeV,
with a tail extending out to about 50~GeV~\cite{Acquistapace:1998rv}.

The highest energy neutrino beam in operation so far has been the
Fermilab wide band neutrino beam used by the CCFR experiment~\cite{McFarland:1995sr}.
It was produced by 800~GeV protons incident on a production target,
with quadrupole magnets focusing a wide band of secondary particles
onto the experiment. The mean energy for the Fermilab wide band beam
was about 80~GeV, but interactions due to neutrinos of up to about
500~GeV were observed.

Based on the discussion above, it is easy to see how high-energy neutrino
beams could be constructed at Fermilab and at Cern. At Fermilab, protons
from the Main Injector are already being used to produce a neutrino
beam, but in order to reach higher energies, the Tevatron needs to
be set up in a new high-energy neutrino beam configuration. At Cern,
the SPS can be used to produce a wide-band neutrino beam (rather than
the lower energy beam that is currently planned for CNGS). But to
reach even higher energies, the highest energy accelerator, the LHC,
needs to be reconfigured to produce neutrino beams. We will propose
specific beam configurations in this paper. 

Since every pion decay produces not only a neutrino but also a partner
muon, these proton-produced neutrino beams have associated muon beams
with energy distribution and beam parameters comparable to the neutrino
beam itself. This is a serious challenge when it comes to the detector
setup, because the muons penetrate the interaction region, the detector,
and even interact with the detector material. It will be important
to separate interactions caused by muons from interactions caused
by neutrinos in the detector. The muons don't all have to be absorbed
before they decay (such as required for a long-baseline neutrino beam),
but they need to be deflected away from the neutrino interaction region.
The associated muon beam will be discussed further in Sec.~{sub:muonbeam}.


\subsubsection{Tevatron\label{sub:TevatronBeam}}

The Fermilab Tevatron has been used in the past to produce a high-energy
neutrino beam from 800~GeV protons incident on a Beryllium target.
The Tevatron currently accelerates protons and antiprotons to 980~GeV
each and then collides them in two interaction regions. It should
be possible to extract the protons at 980~GeV from the Tevatron and
direct them onto a target as shown in Fig.~\ref{fig:FermilabNuBeamMI}.

The number of neutrinos produced by such a beam can be calculated
assuming a similar layout to the current NuMI beamline. The existing
Main Injector neutrino beamline accepts 5 bunches of protons from
the Booster per 0.45~Hz spill, each bunch containing $5\times10^{12}$
protons~\cite{Crane:1995ky}. Work is currently underway to double
this number~\cite{Koba:2003sj}. An increase in the number of protons
by another order of magnitude is currently being studied in the context
of neutrino superbeams. This requires a high-intensity proton driver
to deliver significantly more protons into the accelerator complex.
Thus, we assume as a baseline that the Tevatron can produce a proton
beam containing $10^{14}$~protons per bunch and provide $5\times10^{7}$~bunches
per year. 

The parameters of the resulting neutrino beam have been calculated
with the bmpt program~\cite{Bonesini:2001iz}. For simplicity, the
target parameters have been taken as the default bmpt settings, which
refer to the CNGS neutrino beam~\cite{Acquistapace:1998rv}. Only
the parameters for incident proton energy, pion decay tunnel length,
detector distance, and detector radius have been adjusted. The incident
proton beam energy is set to 980~GeV and the proton target is
made of carbon. 

\begin{figure}
\includegraphics[scale=0.4]{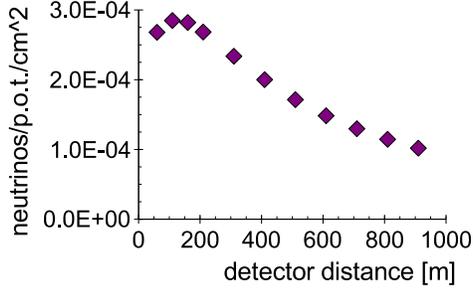}

\caption{Neutrino flux in a circular region with radius 1~cm as a function
of the distance to the proton target, for a proton beam of 980~GeV,
for optimal secondary particle focusing and a decay tunnel that is
10~m shorter than the detector distance. \label{fig:FermilabNuDistance}}

\end{figure}

Fig.~\ref{fig:FermilabNuDistance} shows the neutrino flux in a small
cross section area with 1~cm radius as a function of the distance
to the proton target. The optimal interaction point is at a location
between 100~m and 200~m downstream of the production target. At
shorter distances the decay tunnel is too short for enough pions to
decay, while at larger distances the neutrino beam diverges too much.
The flux maximum is fairly broad however, which makes the placement
of the different components flexible.

\begin{figure}
\includegraphics[scale=0.4]{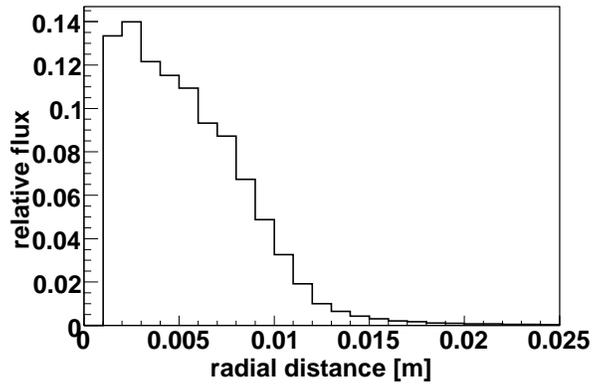}

\caption{Radius of the neutrino beam at a distance of 210~m downstream of
a proton target for the Fermilab Tevatron beam, extracted at 980~GeV.\label{fig:FermilabNuRadial}}

\end{figure}

For the following calculations we choose a 200~m long decay tunnel
which contains two focusing magnets. Downstream of the decay tunnel
is a 10~m thick shielding wall. This space should also contain magnets
to deflect muons and pions and other charged particles away from the
interaction region. The radial profile of the neutrino beam at this
location is shown in Fig.~\ref{fig:FermilabNuRadial}. The initially
point-size beam spreads out to a radius of about 1~cm.

Henceforth, the vertical dimensions of the neutrino interaction region
will be set to a radius of 1~cm. This is a significantly larger beam
size than in typical charged particle colliders, but is unavoidable
because the neutrinos themselves cannot be focused. 

\begin{figure}
\includegraphics[scale=0.4]{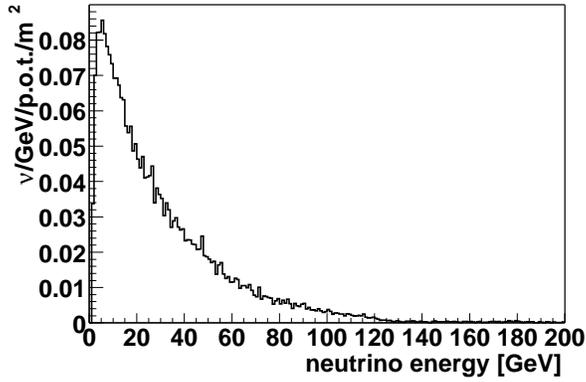}

\caption{Neutrino energy at a distance of 210~m downstream of a proton target
for the Fermilab Tevatron beam, extracted at 980~GeV.\label{fig:FermilabNuEnergy}}

\end{figure}

The neutrino energy spectrum in this configuration is shown in Fig.~\ref{fig:FermilabNuEnergy}.
The spectrum is falling quickly and there are almost no neutrinos
left with energies above 120~GeV. The mean neutrino energy is 30~GeV.

\subsubsection{LHC}

\begin{figure}
\includegraphics[scale=0.4]{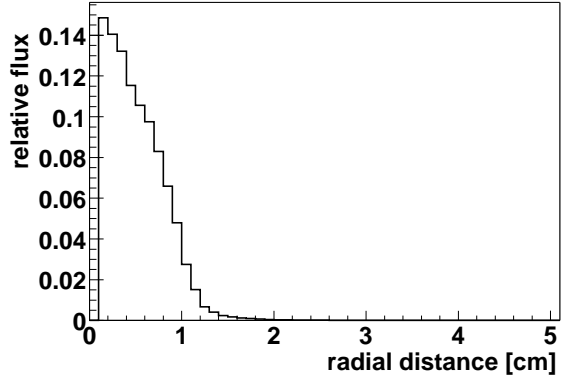}

\caption{Radial profile of the neutrino beam at a distance of 210~m downstream
of a proton target in the LHC configuration.\label{fig:LHCNuRadial}}

\end{figure}

The LHC will accelerate protons to 7000~GeV and collide those in
four separate interaction regions. Using these protons to produce
neutrinos will result in a high-intensity, high-energy neutrino beam.
The radial distribution of such a neutrino beam at a distance of 210~m
is shown in Fig.~\ref{fig:LHCNuRadial}. The Gaussian width is about
0.5~cm. The beam is more focused than the Tevatron neutrino beam
due to the higher incident proton energy.

\begin{figure}
\includegraphics[scale=0.4]{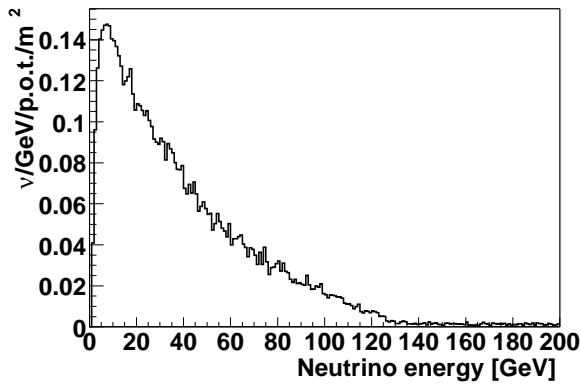}

\caption{Neutrino energy at a distance of 210~m downstream of a proton target
for the LHC beam, extracted at 7000~GeV.\label{fig:LHCNuEnergy}}

\end{figure}

The neutrino energy distribution resulting from a 7000~GeV proton
beam is shown in Fig.~\ref{fig:LHCNuEnergy}, again using the BMPT
program with a detector at a distance of 210~m. The neutrino energy
spectrum is very similar to the Tevatron, but shifted to higher energies,
with a mean energy of about 40~GeV.

\subsection{Neutrinos from muon beams}

If muons are accelerated to very high energies, then their decay $\mu\rightarrow e\overline{\nu}_{e}\nu_{\mu}$
produces a high energy, high intensity neutrino beam. Accelerating
muons rather than protons results in significantly more neutrinos
at significantly higher energies. Since the neutrinos originate from
the muon decay, a large fraction of the initial beam energy is transferred
to the neutrinos and much higher neutrino energies can be accomplished
with muon beams than with proton beams of comparable energy. Conversely,
muon beams don't have to be accelerated to as high of an energy in
order to obtain similar neutrino beam energies as obtained by a proton
beam. There has been a lot of excitement in recent years about muon
accelerators and their potential in producing neutrino beams. Neutrino
factories are envisioned that utilize muon storage rings to produce
high-intensity neutrino beams~\cite{Geer:1997iz,nufact}. A detailed
study of the Physics potential of a neutrino superbeam produced by
muons was carried out~\cite{Albright:2000xi}. Much of the infrastructure
required for this neutrino beam is also needed for a muon collider,
and one facility might be able to provide both. As will be seen below,
a neutrino superbeam can be used not only in fixed-target experiments
but also in a collider configuration. 

The muons are produced from a proton beam incident on a stationary
target. They are then collected, accelerated, and put into the storage
ring which has a long straight section. Muon decays in the straight
section result in an intense and well-focused neutrino beam. 

The beam intensity for such neutrino beams is significantly higher
than those discussed above due to the acceleration of the muons. Current
estimates give a yield of about 0.1~to 0.2~muons per proton~\cite{Albright:2004iw}.
The same proton drivers as mentioned above thus yields approximately
$10^{13}$ muons per bunch in the accelerator. About a third of them
decay in the straight section. The resulting neutrino beam is tightly
focused, much better than from proton machines due to the better focusing
of the decay particles and the Lorentz boost. Most of the neutrinos
pass through an interaction region with a radius of about 1~mm located
immediately downstream of the straight section. Such a configuration
yields about $10^{12}$~neutrinos per bunch of $10^{14}$~protons.
The energy distribution of the neutrino beam from such an arrangement
are given in Ref.~\cite{Geer:1997iz}. Due to the muon decay kinematics,
the muon neutrino energy peaks close to the muon beam energy for neutrinos
that travel within a small angle of the original muon beam.

\subsection{Neutrinos from pion beams}

If a muon collider is not the ultimate goal, then accelerating pions
has significant advantages over accelerating muons. Pions are by far
the most abundant particle produced when a proton beam hits a beam
dump. Thus, it might be easier to construct a pion accelerator than
a muon accelerator. At the same time, many of technical challenges
are common to pion and muon accelerators, in particular the short
lifetime of the accelerated particles. Thus, a pion accelerator could
provide a first step towards a muon accelerator. Similar to muon beams,
neutrinos are produced in the decay of pions, and hence one could
construct neutrino superbeams based on pion beams. However, no such
accelerator is currently being planned, and neutrinos from pion beams
will not be discussed further in this paper.

\section{Colliding neutrinos}

Since neutrino beams are produced by accelerating charged particles
rather than the neutrinos themselves, any neutrino beam facility will
also have associated beams of charged particles. It is possible to
utilize these in conjunction with the neutrino beam. This leads to
two basic colliding neutrino beam configurations. 

\begin{enumerate}
\item Colliding neutrinos with other particles such as protons and electrons
or even pions and muons. As discussed in Section~\ref{sec:Neutrino-Beams},
the associated beams are not only needed but also have higher energy
than the actual neutrino beam. One exception is an electron beam which
would need to be constructed separately.
\item Colliding neutrinos with neutrinos. This configuration could involve
neutrinos of the same flavor or of a different flavor. The neutrino
beams discussed below all contain a mixture of flavors.
\end{enumerate}
The rate of neutrino interactions $R_{event}$ in a colliding beam
setup is calculated from the cross section $\sigma$ for a given process
as

\[
R_{event}={\cal L}\sigma,\]

where the luminosity $L$ is given by 

\begin{equation}
{\cal L}=\frac{N_{1}N_{2}}{4\pi\sigma_{x}\sigma_{y}}.\label{eq:lumi}\end{equation}
 Here, $N_{1}$w and $N_{2}$ are the number of particles in the two
beams, and $\sigma_{x}$ and $\sigma_{y}$ are the Gaussian widths
of the interaction region in the plane perpendicular to the beam direction~\cite{PDBook}. 

In the following, we use the Madgraph program to estimate cross sections
for various processes~\cite{Maltoni:2002qb}. The neutrino energy
spectra from Figs.~\ref{fig:FermilabNuEnergy} and~\ref{fig:LHCNuEnergy}
have been added to Madgraph in parametrized form. The neutrino energy
distribution from a muon collider are based on Ref.~\cite{Geer:1997iz}
and have also been added to Madgraph in parametrized form.

\subsection{Colliding neutrinos with protons}

We assume that the same proton beam that produces the neutrinos is
also used in neutrino-proton collisions.

\subsubsection{Tevatron}

\begin{figure}
\includegraphics[scale=0.55]{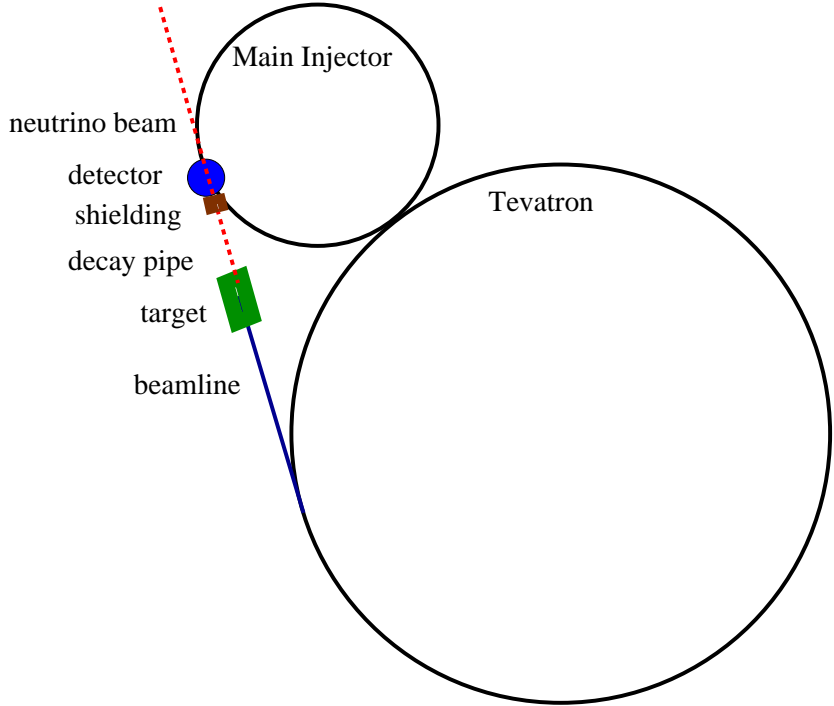}

\caption{Layout of the Fermilab accelerator complex as a neutrino collider,
extracting protons from the Tevatron and directing them onto a target
such that the resulting neutrino beam collides with the proton beam
circulating in the Main Injector.\label{fig:FermilabNuBeamMI}}

\end{figure}

The Fermilab accelerator complex can be set up to collide neutrinos
with protons circulating in the Main Injector in a straightforward
way. Fig.~\ref{fig:FermilabNuBeamMI} shows this arrangement, which
uses the same proton beam in the collisions and the neutrino production.
Using the same parameters as in Sec.~\ref{sub:TevatronBeam}, the
proton beam contains $10^{14}$ protons per bunch. However, it has
a significantly smaller cross section area than the neutrino beam.
This results in an instantaneous luminosity of $4\times10^{23}\, cm^{-2}$
for each collision of a neutrino bunch with a proton bunch. Assuming
$5\times10^{7}$~bunches per year, and that each neutrino bunch collides
with one proton bunch, yields an integrated luminosity of $20\,\mu b^{-1}$
per year. The neutrino-proton weak interaction cross section for the
parameters given above is $1.5\times10^{-3}\,\mu b$, calculated with
the Madevent program~\cite{Maltoni:2002qb}. Hence at this luminosity
there are about 0.03 neutrino-proton interactions per year, or 3 interactions
every 100 years.

Reaching luminosities required for reasonable event samples on the
order of several $\textrm{pb}^{-1}$ (for example comparable to the
startup of the HERA electron-proton collider~\cite{Ahmed:1995fd,Derrick:1994sz})
requires an increase in the proton intensity by about two orders of
magnitude to $10^{16}$ protons per bunch. Such intensities result
in a few thousand neutrino-quark charged current scattering events
per year, sufficient for initial studies. Over time, the intensity
will need to be increased even further to reach much higher integrated
luminosities required to start detailed studies of neutrino-proton
interactions at the highest energies.

Alternatively, if a muon storage ring is built at Fermilab, then neutrinos
from this beam can be directed onto the Tevatron proton beam. According
to Eq.~\ref{eq:lumi}, with two beams that are each about 1~mm in
size, an instantaneous luminosity of about $8\times10^{26}cm^{-2}$
can be reached. Assuming that such an accelerator setup could also
collide $5\times10^{7}$ bunches each year gives an integrated luminosity
of $4\, pb^{-1}$ per year. This is a factor 2000 higher than the
setup with a Tevatron-based neutrino beam described above, mainly
due to the significantly smaller transverse size of the interaction
region. This will be sufficient for initial studies. If the intensity
of both the proton and muon beams is increased by another order of
magnitude then several hundred $pb^{-1}$ can be collected each year
allowing for high-statistics physics studies.

\subsubsection{LHC}

If the LHC is configured to direct the 7000~GeV proton beam onto
a production target then the resulting neutrino beam can be directed
at the SPS accelerator where the neutrinos would collide with 450~GeV
protons. This configuration is equivalent to the Fermilab Tevatron-Main
Injector setup, but the corresponding energies are higher. The neutrino-proton
cross section increases to $6.9\times10^{-3}\,\mu b$, where the calculation
again uses the Madgraph program and the neutrino energy spectrum shown
in Fig.~\ref{fig:LHCNuEnergy}. Since the LHC hasn't started running
yet and will take quite a few years to complete its physics program,
reaching proton beam intensities of $10^{16}$ protons per bunch might
become realistic by the time the neutrino beam could be built.

\subsection{Colliding neutrinos with leptons}

\subsubsection{Tevatron}

In order to collide the neutrino beam with particles other than protons
at the Tevatron or LHC, these particles will need to be produced and
accelerated separately. They cannot be part of the proton acceleration
process. At the Tevatron, there is already a second accelerator located
in the same tunnel as the Main Injector, and that is the Recycler.
If the Recycler can be converted to accelerate electrons rather than
storing antiprotons, then it becomes possible to collide neutrinos
produced by the Tevatron beam with electrons. The electron beam wouldn't
be of very high energy, but energies above 20~GeV might be achievable
and should already be sufficient for interesting physics studies.

The resonance production of $W$~bosons from neutrinos and electrons
according to Fig.~\ref{fig:FeynmanNeuLep}~(c) has a cross section
of the order of $1pb$ at such a collider. The challenge is thus to
produce a high intensity electron beam in order to obtain reasonable
interaction rates. For example, in order to reach luminosities of
several ${pb}^{-1}$, the beam needs to contain more than $10^{18}$
electrons per bunch, which is about seven orders of magnitude higher
than what has been accomplished in past electron-positron colliders~\cite{PDBook}.
Producing such a very-high intensity electron beam requires significant
technological advances. Thus neutrino-electron colliders are not likely
to be feasible in the near future, even if the intense neutrino beams
become a reality.

\subsubsection{Muon storage ring}

\begin{figure}
\includegraphics[scale=0.55]{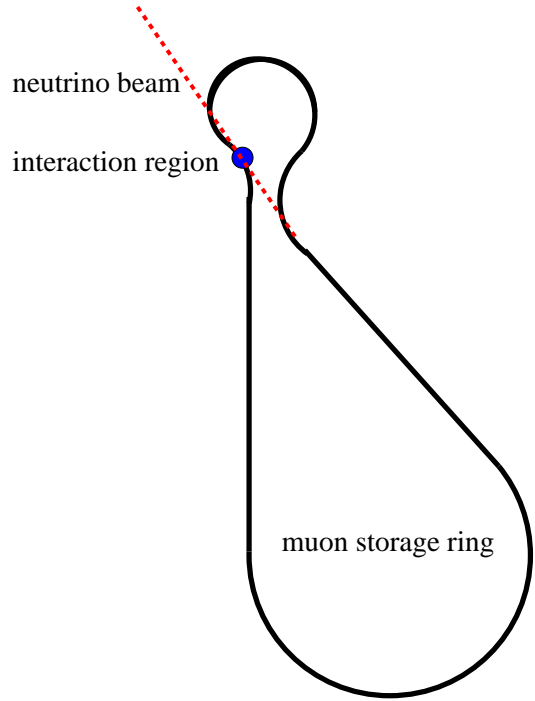}

\caption{Neutrino-lepton collider based on a muon storage ring. There is only
one muon beam circulating in the storage ring. The neutrino-muon interaction
region is indicated by the small circles.\label{fig:muonRingNu}}

\end{figure}

A neutrino-muon collider based on a muon storage ring solves this
problem. It can provide much higher CM energy and luminosity. It provides
a both the high-energy muon beam and the neutrino beam. An example
layout that works with a single muon beam is shown in Fig.~\ref{fig:muonRingNu}.

\begin{figure}
\includegraphics[scale=0.55]{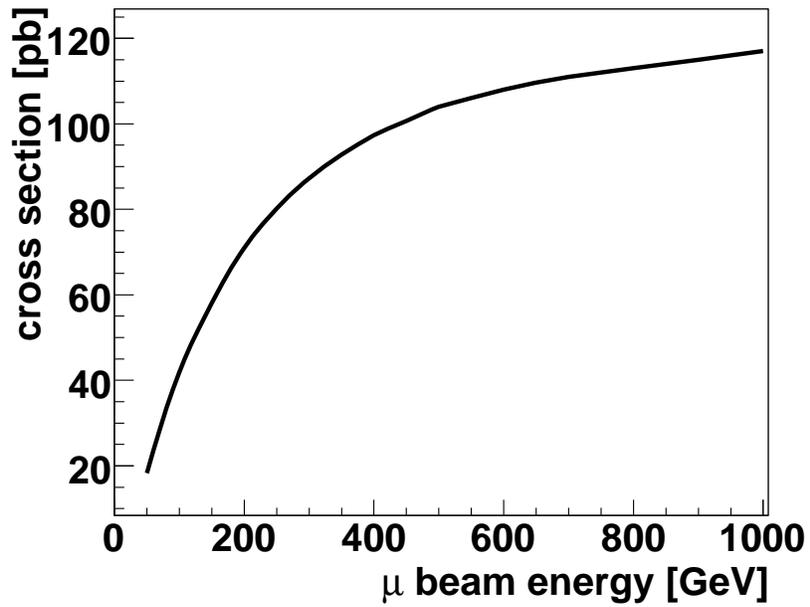}

\caption{SM cross section for the process $\nu_{\mu}\mu\rightarrow\nu_{\mu}\mu$
in a neutrino-muon collider, as a function of muon beam energy.\label{fig:numumxs}}

\end{figure}

In this configuration, the resonance production of $W$~bosons is
not allowed because of the $\mu\nu_{\mu}$ initial state. The main
SM process is the t-channel exchange of a $W$~boson, see Fig.~\ref{fig:FeynmanNeuLep}~(b).
The cross section for this process is shown in Fig.~\ref{fig:numumxs}.

Since the neutrinos were produced from the same beam they are colliding
with and hence have the same flavor, annihilation into a $W$~boson
is not possible, only $t$-channel exchange of a $W$~boson. This
is reflected in the relatively low cross section. The distribution
is shown as a function of beam energy rather than CM energy because
in contrast to the muon, the energy of the initial state neutrino
is not known.

\begin{figure}
\includegraphics[scale=0.55]{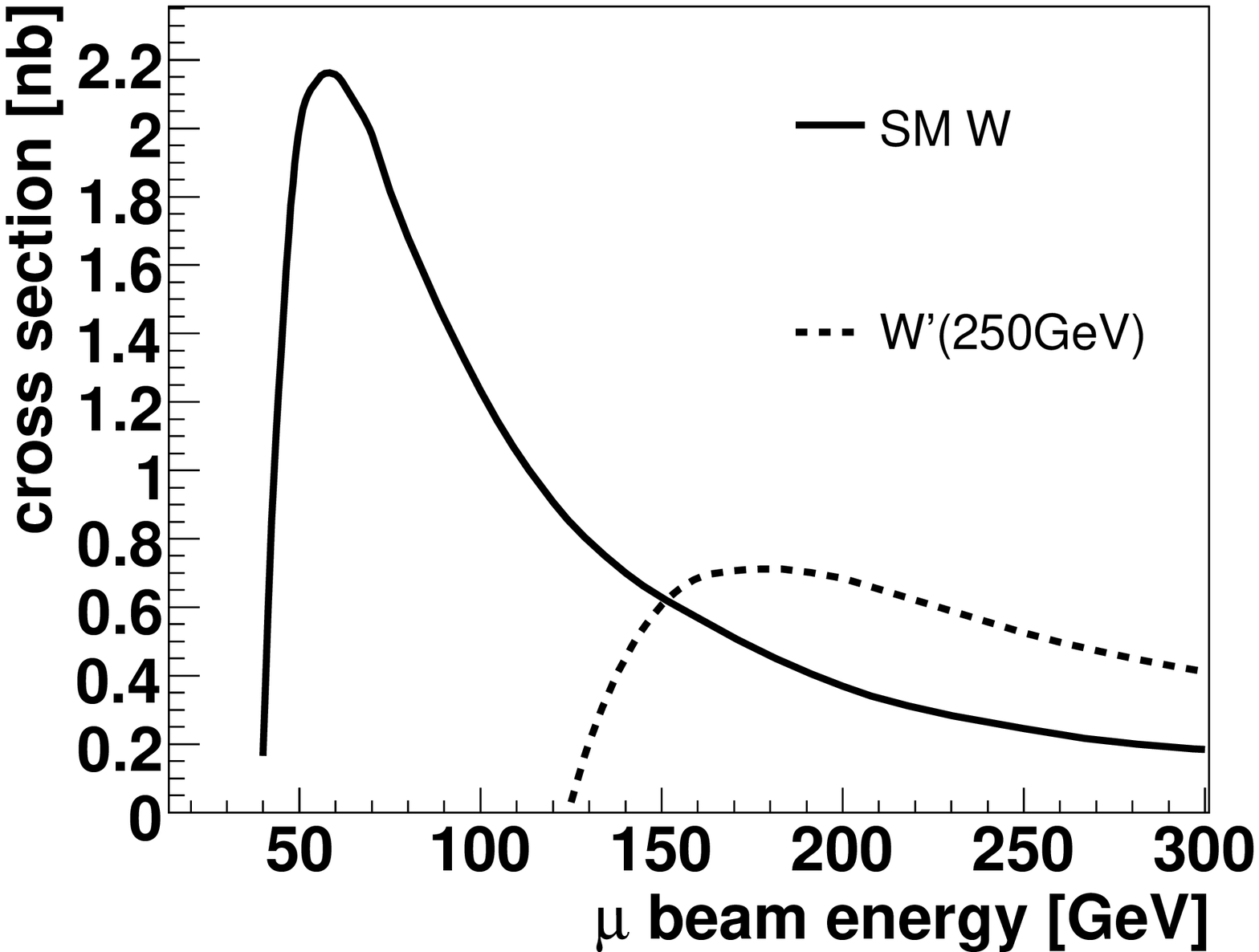}

\caption{Cross section for neutrino-muon annihilation into a SM $W$~ boson
or a heavy $W'$~boson in a neutrino-muon collider.\label{fig:numupxs}}

\end{figure}

If instead two muon beams are used, as shown in Fig.~\ref{fig:muonColliderb}
below, then neutrinos collide with anti-muons or vice versa, making
resonance $W$~boson production possible. The cross section for this
process is shown in Fig.~\ref{fig:numupxs}.

The cross section for this process as a function of the muon beam
energy is shown as the solid line in Fig.~\ref{fig:numupxs}. While
there is no sharp resonance peak due to the spread in neutrino energies,
the cross section nevertheless shows a clear peak. 

If there is new physics involving an additional heavy charged boson
(such as the $H^{+}$ shown in Fig.~\ref{fig:FeynmanNeuLepNP}
or a heavy $W'$~boson), then this can also be produced in resonance.
If this boson couples only to leptons and neutrinos and not to quarks,
then it will not be produced at hadron or lepton colliders. In this
case a neutrino-lepton collider is the best place to produce such
a heavy boson and measure its properties. The dashed line in Fig.~\ref{fig:numupxs}
shows the cross section for $W'$~boson production with a mass of
250~GeV.

\begin{figure}
\includegraphics[scale=0.55]{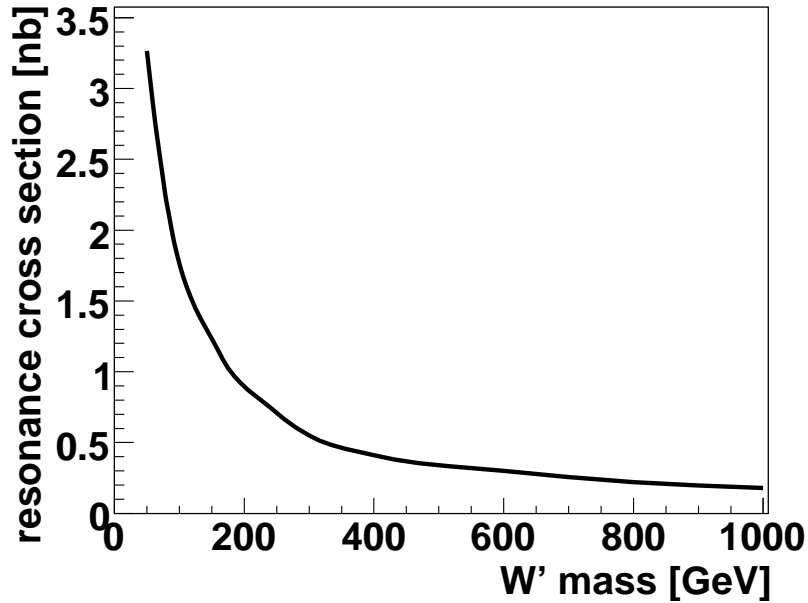}

\caption{Peak cross section for neutrino-muon annihilation into a heavy $W'$~boson
in a neutrino-muon collider as a function of the $W'$~boson mass.\label{fig:numupwpresxs}}

\end{figure}

The peak cross section occurs at a beam energy corresponding to about
75\% of the boson mass. The dependence of this peak cross section
on the mass of the $W'$~boson is shown in Fig.~\ref{fig:numupwpresxs}.
The peak cross section is above 1~nb for $W'$~boson masses up to
about 200~GeV and remains above 0.3~nb for masses as high as 1~TeV.

The event rate in a neutrino-muon collider is already quite large
with the planned proton driver discussed above. Assuming that in each
collision, $10^{12}$ neutrinos from a $\mu^{+}$ beam collide with
a $\mu^{-}$ beam containing $10^{13}$ muons, then his yields an
instantaneous luminosity of about $10^{27}\, cm^{-2}$ per bunch crossing.
Current plans for a muon collider assume running at 15~Hz with four
bunches. That gives a luminosity of $6\times10^{28}\, cm^{-2}s^{-1}$
which is equivalent to a yearly ($10^{7}$~s/year) integrated luminosity
of $600\, nb^{-1}$. This is sufficient to produce both SM $W$~bosons
and new $W'$~bosons or any other new particle with similar couplings.

\subsection{Colliding neutrinos and neutrinos}

\subsubsection{LHC}

\begin{figure}
\includegraphics[scale=0.55]{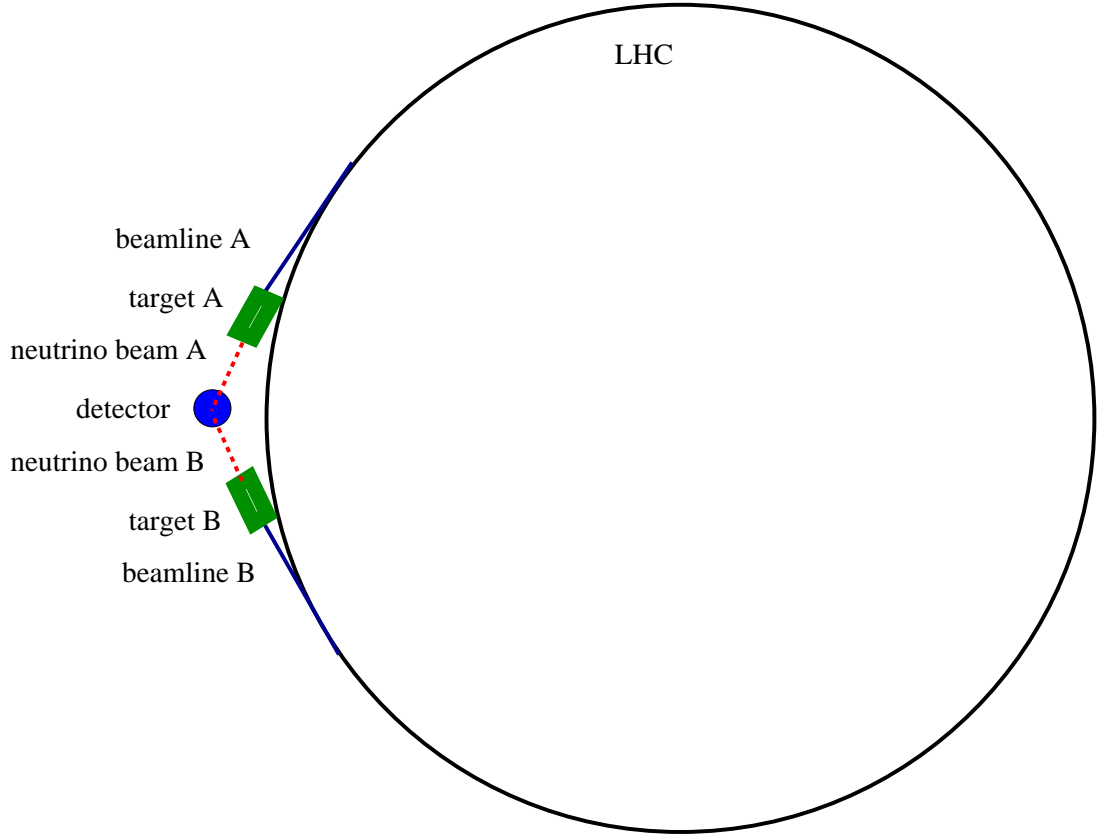}

\caption{Layout of the LHC accelerator as a neutrino collider, extracting both
proton beams and directing them onto two targets such that the resulting
neutrino beams collide with each other.\label{fig:LHCNuCollide}}

\end{figure}

The LHC will accelerate protons to 7000~GeV and collide those in
four separate interaction regions. Since the LHC collides a proton
beam with another proton beam, both of them can be used to produce
neutrino beams and the LHC can become the first neutrino-neutrino
collider. This is not possible at the Tevatron where one of the two
beams contains antiprotons, and significantly fewer of them than protons.
In order to set up a neutrino collider, both proton beams are extracted
from the LHC at 7000~GeV and directed onto two different production
targets. Pions and kaons are allowed to decay in two separate decay
tunnels. The resulting neutrino beams intersect each other at a central
detector location. The remaining pions, muons, and other secondaries
need to be absorbed or diverted from the neutrino direction through
shielding and magnetic fields. The layout of such a machine is shown
in Fig.~\ref{fig:LHCNuCollide}.

Assuming a similar proton intensity per bunch can be achieved at the
Tevatron and at the LHC($10^{14}$~protons per bunch), then the luminosity
per crossing is about $2\times10^{24}\, cm^{-2}$. Since there are
now two beams that have to be fed from the accelerator chain, the
repetition rate will likely be smaller than at the Tevatron and we
thus assume about $10^{7}$ collisions per year. This gives an integrated
yearly luminosity of $20\mu b^{-1}$. The cross section for the process
$\nu\overline{\nu}\rightarrow\ell\overline{\ell}$ is about 0.2~nb,
thus a setup like this produces about 0.004 di-lepton events per year.
In order to produce sizable samples of several tens of events per
year, the intensity of the proton beams needs to be increased by about
two orders of magnitude.

Similarly to the Tevatron case, in order to achieve integrated luminosities
of several $pb^{-1}$ which makes it possible to study more physics
processes, the intensity of the proton beams will need to be increased
by about three orders of magnitude. While there is currently no clear
path to accomplish this, the LHC will run as a proton-proton collider
for many years. That time can be spent on increasing the proton luminosity,
both at the Tevatron and at the LHC. By the time the LHC will be finished
running and could possibly be converted to a neutrino collider, significantly
higher intensities are likely achievable.

\subsubsection{Muon collider}

\begin{figure}
\includegraphics[scale=0.55]{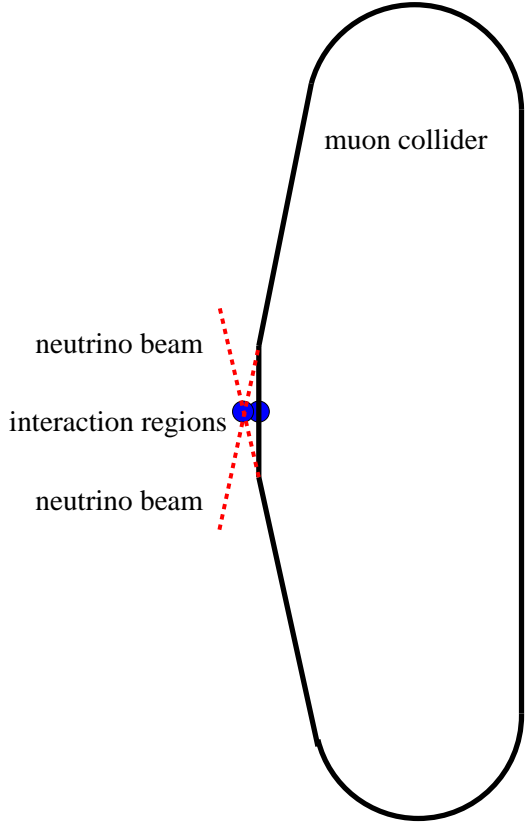}

\caption{Muon beam configuration to use a muon collider also as a neutrino
collider. Each of the two counter-circulating muon beams produces
a high-intensity neutrino beam. The two neutrino beams collide with
each other in close proximity to the muon collision point. The two
interaction regions are indicated by the small circles.\label{fig:muonCollidera}}

\end{figure}

A muon collider consists of two counter-circulating beams that are
brought into collision at one or two interaction points~\cite{Finley:1999yt,Ahn:1999kd}.
Each of the muon beams produces a high-intensity, narrowly focused
beam of neutrinos, comparable to the neutrino factory. The layout
of the two muon beams in the collider has to be changed only slightly
in order to achieve collisions of the two neutrino beams. An simple
layout is shown in Fig.~\ref{fig:muonCollidera}. One of the straight
sections of the muon collider is divided into two half-length sections,
which have a small angle with respect to each other. Very close to
the interaction point, the muon beams are then bent towards each other.
The neutrino beams resulting from muon decays in the straight sections
are not deflected and continue on a straight path. The neutrino collision
point is thus shifted a small distance away from the muon beam direction.

If the angles are chosen such that the two straight sections are almost
parallel, then the neutrino interaction region is so close to the
muon interaction region that it might be possible to enclose both
interaction regions within the same detector. This allows for significant
savings in detector cost because only one detector needs to be built
rather than two.

\begin{figure}
\includegraphics[scale=0.55]{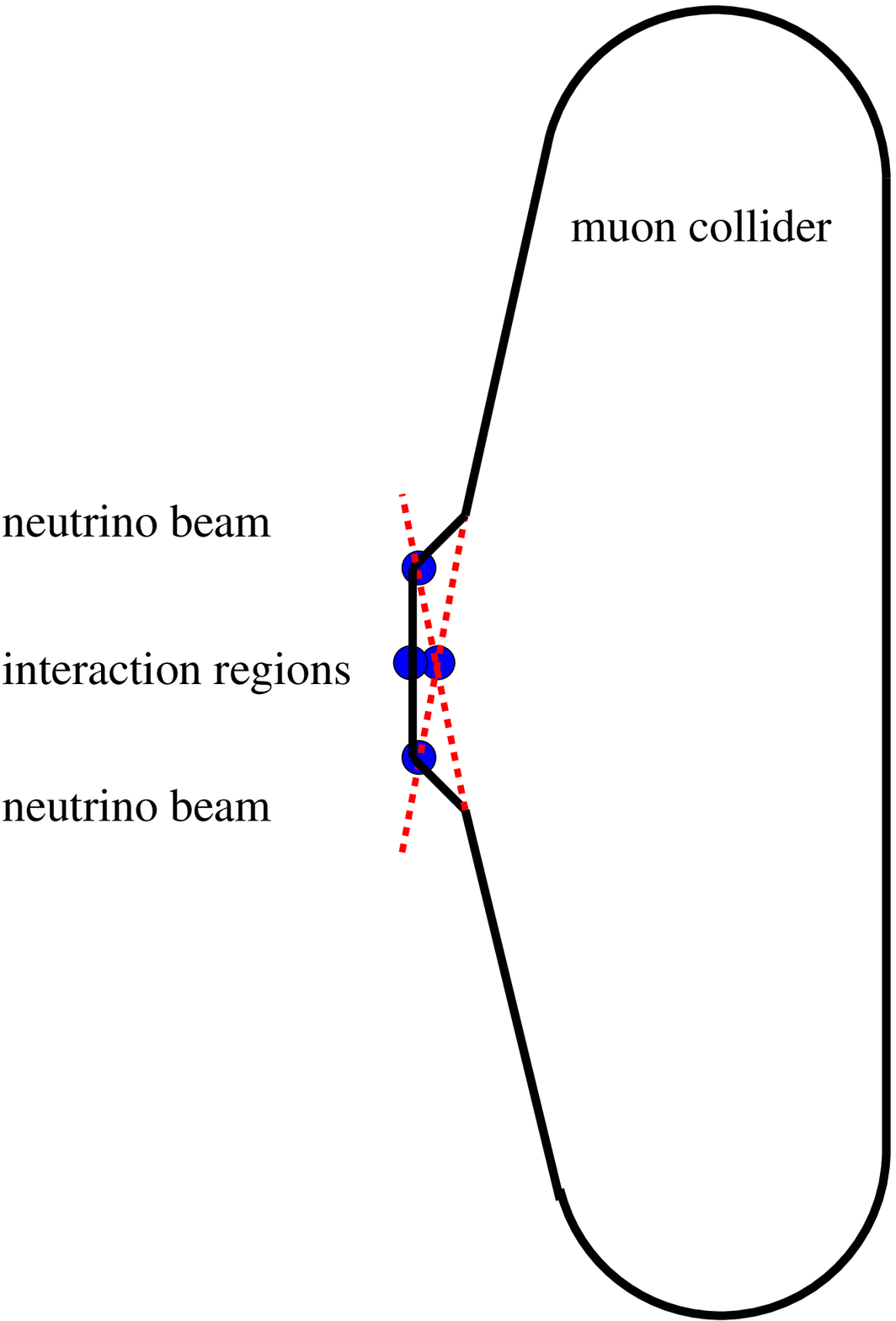}

\caption{Muon beam configuration to use a muon-muon collider also as a neutrino-neutrino
and neutrino-muon collider. Each of the counter-circulating muon beams
produces a high-intensity neutrino beam. The two neutrino beams collide
with each other and with the muon beam. The four interaction regions
are shown as small circles.\label{fig:muonColliderb}}

\end{figure}

An alternative colliding beam setup that provides not one or two but
four interaction regions is shown in Fig.~\ref{fig:muonColliderb}.
In this configuration, the neutrino beams not only collide with each
other but also with the muon beams themselves. The four interaction
regions are in close proximity to each other: a muon-muon interaction
region, a neutrino-neutrino interaction region, and two neutrino-muon
interaction regions.

Since the muon collider configuration relies on the same basic storage
ring parameters as the neutrino factory, we can use the same parameters
as above to calculate the luminosity for this configuration. We assume
that in each collision, $10^{12}$ neutrinos from a $\mu^{+}$ beam
collide with $10^{12}$ neutrinos from a $\mu^{-}$ beam in an interaction
region with a radius of about 1~mm. This yields an instantaneous
luminosity of about $10^{26}\, cm^{-2}$ per bunch crossing. Current
plans for a muon collider assume running at 15~Hz with four bunches.
That gives a luminosity of $6\times10^{27}\, cm^{-2}s^{-1}$ which
is equivalent to a yearly ($10^{7}$~s/year) integrated luminosity
of $60\, nb^{-1}$. This is a factor 3000 higher than what could be
achieved by converting the LHC to a neutrino collider. Hence, this
setup is already sufficient to observe $Z$~boson production in neutrino-anti-neutrino
annihilation. Moreover, the neutrino energy in this configuration
is significantly higher (a multi-TeV muon collider will produce multi-TeV
neutrinos) than in the LHC-based neutrino collider. It will for example
be sensitive to higher-mass resonances from new physics in neutrino
annihilation.

\begin{figure}
\includegraphics[scale=0.55]{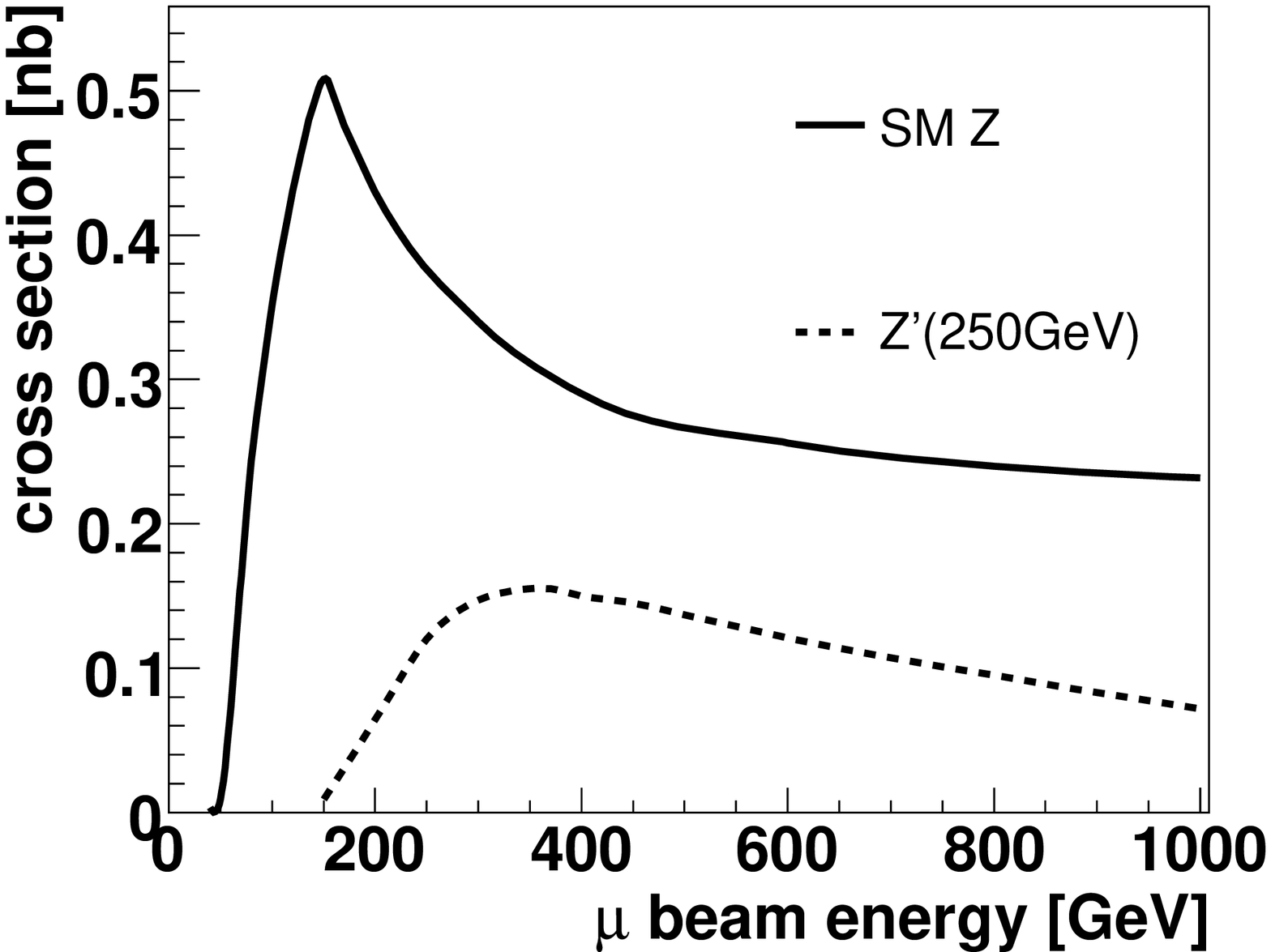}

\caption{Cross section for neutrino-neutrino scattering to a lepton-lepton
final state at a muon collider, as a function of the muon beam energy
(solid line). Also shown is the cross section for the production of
a heavy $Z'$~boson.\label{fig:vmvmzp}}

\end{figure}

Fig.~\ref{fig:vmvmzp} shows the cross section for the production
of a $Z$~boson at the muon-collider based neutrino collider, as
a function of the muon beam energy. The cross section shows a clear
peak at a muon energy of slightly less than twice the $Z$~boson
mass, similar to $W$~boson production shown in Fig.~\ref{fig:numupxs}.
Also shown is the cross section for production of a heavy $Z'$~boson,
with neutrino couplings identical to the SM $Z$~boson but at a mass
of 250~GeV. In this case the resonance peak is significantly broader,
but the cross section is large enough to be able to observe the $Z'$~boson.

Furthermore, since the muon collider is currently only in its planning
stages, there is still much room for further improvements. In particular,
if the intensity can be increased by another order of magnitude, then
integrated luminosities of several $pb^{-1}$ can be reached. Such
luminosities extend the sensitivity to new physics in the neutrino
sector significantly. 

\begin{figure}
\includegraphics[scale=0.55]{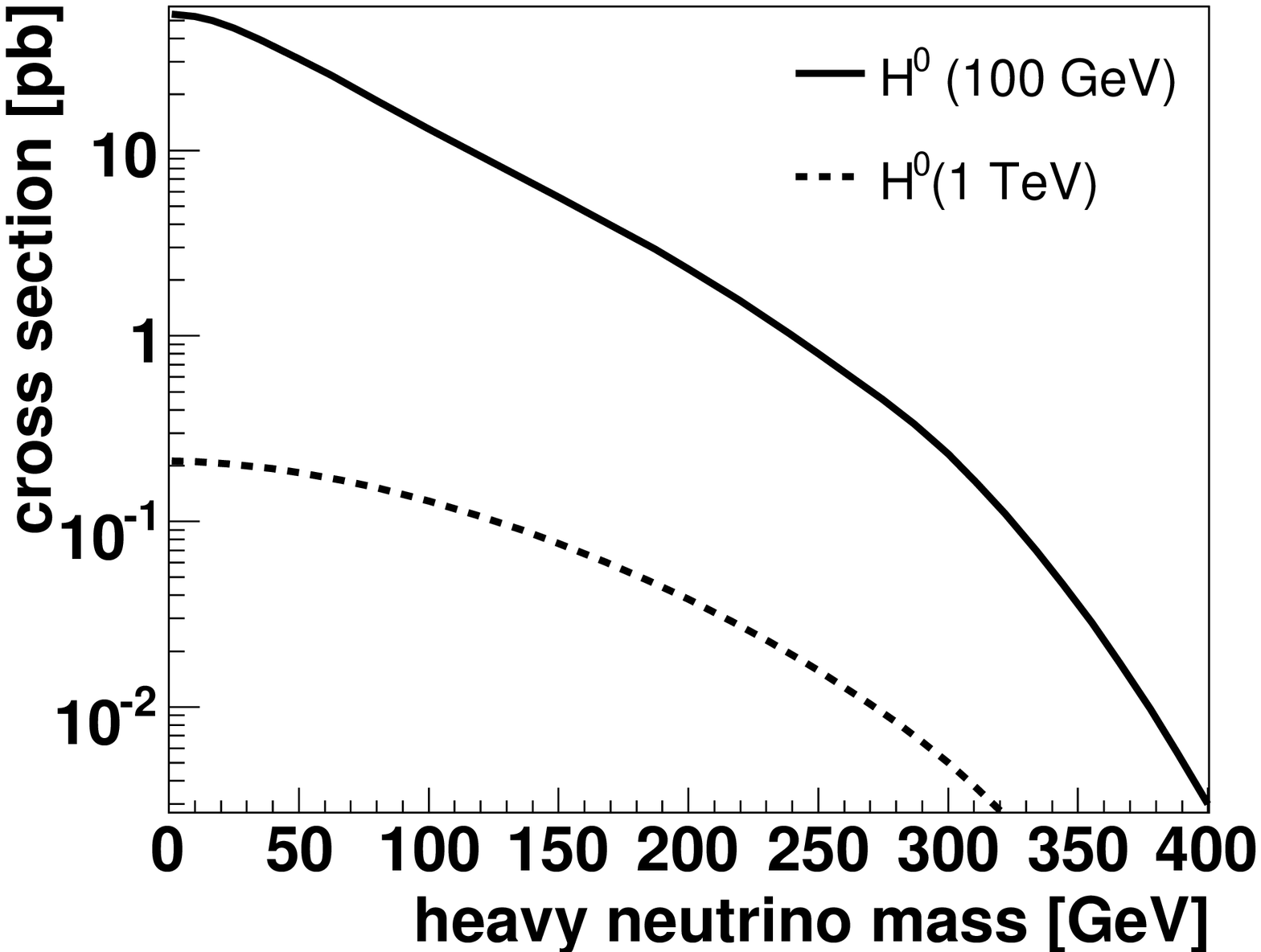}

\caption{Heavy neutrino pair production cross section in neutrino-neutrino
scattering, as a function of the heavy neutrino mass, for two different
Higgs boson masses. The neutrinos beams are produced by a muon collider
with 500~GeV muon beam energy and the $\nu NH^{0}$ coupling strength
is 1.\label{fig:vmvmNN}}

\end{figure}

For example, Fig.~\ref{fig:vmvmNN} shows the cross section for the
production of heavy neutrinos according to Fig.~\ref{fig:FeynmanNeuNeuNP}~(a).
A $\nu NH^{0}$ coupling strength of 1 is assumed in this example,
the exact coupling strength depends on the specific model and might
be lower. The neutrinos are produced by 500~GeV muon beams. The cross
section in this scenario is in the pb range for heavy neutrinos up
to a mass of about 250~GeV, for a Higgs boson mass of 100~GeV. If
the Higgs boson is heavier (1~TeV or more), the cross section
is in the fb range and significantly higher luminosity is needed.
Single Higgs boson production through loop processes such as shown in
Fig.~\ref{fig:vmvmNN}~(c) will also be relevant, but the cross section 
calculation is beyond the scope of this paper.

\section{Common Experimental Challenges}

While neutrino colliders provide exciting measurement opportunities,
they also pose unique experimental challenges. As far as the neutrino
beam is concerned, there are two main challenges. First and foremost,
in order to achieve the required luminosities at a proton accelerator,
it is necessary to increase the intensity of the primary proton beams
beyond what is currently achievable by several orders of magnitude.
Second, experiments will have to deal with many interactions in the
detector, due to both beam neutrinos and muons associated with the
neutrino beam.

\subsection{Associated muon beam}
\label{sub:muonbeam}

In a proton beam dump experiment there is always an intense muon beam
from the decay of pion secondaries associated with the neutrino beam.
Since neutrinos possess no charge and cannot be focused, it is imperative
for the collision point to be as close as possible to the neutrino
production point. This makes it a challenging task to keep muons and
other beam-related particles out of the interaction region. In this
paper the distance of 10~m has been used to deflect and absorb non-neutrino
beam particles. However, as Fig.~\ref{fig:FermilabNuDistance} shows,
the neutrino flux does not depend on the distance very strongly and
more space can be reserved to deflect and absorb particles
without reducing intensity significantly.

The intense muon flux can be separated from the neutrino beam with
the help of strong magnetic dipoles to deflect charged particles.
This requires powerful magnets upstream of the interaction region.
Charged particles other than muons similarly need to be deflected
to avoid producing additional showers of hadrons or leptons. 
The highest energy particles will likely not be deflected very far and it 
might be necessary to eliminate some detector material on the two sides to 
which the charged particles are deflected.

Neutral
mesons and baryons can be absorbed by sufficiently think shielding
material in the path of the pion beam. However, any muons passing
through magnets or shielding material produces electromagnetic showers.
Thus, magnets have to be arranged in such a way that the muons pass
through as little material as possible. Minimizing the material in
the muon path is also important to avoid muons scattering into the
interaction region or detector elements. Hence, a balance has to be
found between shielding material and empty space in order to minimize
the non-neutrino flux in the interaction region. The feasibility of
such a deflecting-magnet setup has been established by the DONUT 
collaboration~\cite{donutDiscovery}. 

The situation is different from long-baseline neutrino 
experiments~\cite{Ables:1995wq,Acquistapace:1998rv},
where the associated muon beam is stopped in material in order to
prevent it from reaching a near detector. Instead of stopping the
muon beam, the goal here is to deflect it through powerful magnets
so that it passes by the interaction region and the detector.

\subsection{Interaction region}

The neutrino interaction region is the region where the two neutrino beams 
collide with each other or where the neutrino beam collides with a hadron 
or lepton beam. When the proton beam hits the production target, the 
resulting secondary pion 
beam acquires a large momentum spread and expands in the transverse direction. 
This spread in pion momentum will be reflected in a spread in neutrino momentum 
and divergence of the neutrino beam, as shown in Fig.~\ref{fig:FermilabNuRadial}. 
Hence, the neutrino interaction region will have large transverse dimensions 
compared to other colliding beam experiments.

One advantage of a neutrino-neutrino collider is that it at least in 
principle doesn't require a vacuum beam pipe at the center and also has 
no need for magnetic or electric fields close to the interaction region. 
While the entire detector can at least in principle be instrumented to give 
true $4\pi$ spherical coverage around the collision point, neutrino 
interactions in the detector make this rather impossible.

There will be many interactions between the neutrino beams and the detector 
material itself. The neutrino beam is not narrowly focused and there are many 
neutrinos away from the beam axis. These are mostly at lower energy and thus 
have a smaller interaction cross section. Nevertheless, their interactions with 
the detector and shielding material leads to a large number of energy deposits 
and hits in the sensitive detector elements. The amount of material in any part 
of the detector that is exposed to the neutrino beam will need to be minimized 
in order to reduce these interactions. The path that the neutrinos take will need 
to be as free of material as possible. Indeed, a large vacuum pipe encompassing 
the entire neutrino beam is likely needed in order to eliminate neutrino-detector 
interactions as much as possible.

The transverse dimensions of the neutrino beam are less of a problem in the case of 
a muon storage ring based neutrino collider. Since the muons have uniform momentum 
and very little transverse spread, the neutrinos will also be narrowly 
focussed~\cite{Geer:1997iz}.

\subsection{Detector considerations}

The additional interactions in the detector due to beam neutrinos,
muons, and other beam-associated particles will likely be the largest
source of background to any neutrino collision events. This background
needs to be understood in detail and rejected with high efficiency
in order to be able to observe neutrino-neutrino collisions. On the
negative side, this places serious constraints on possible detector
configurations. On the positive side, these neutrino-detector interactions
can be utilized to measure the neutrino spectrum, intensity, and beam
composition. Close to the collision point, the charged particle tracks
from these beam-detector interactions overlap with the charged particle
tracks from beam-beam collisions and many detectors will have hits
from both types of interactions. Hence, it will be quite challenging
for standard particle detectors as they are in use now to work in
this environment. Far away from the collision point this is less of
a problem because there will be very little neutrino flux. In these
regions standard particle detectors should work just fine, for example
for calorimetry and muon identification.

Separating the charged particle tracks from neutrino collision from
those produced in the detector requires new ideas for sensitive detector
elements. There are two main differences between the particles produced
in neutrino collisions in the interaction region and the beam-related
interactions in the detector that can be exploited here. The first
difference is that beam-related interactions typically result in particles
traveling in the direction of the beam, whereas neutrino collisions
produce particles perpendicular to the beam. The second difference
is that the particles produced in collisions  typically have higher
energy, while particles produced in beam-related interactions with
the detector typically have lower energy. 

Exploiting these differences has the following consequences for possible
detector configurations: Each detector element should be able to distinguish
between different particles traversing it in different directions.
One example of a detector that has this capability is a Cherenkov
detector. An example of this type of detector is the huge water-Cherenkov
detector used in the Kamiokande experiment~\cite{Fukuda:1998mi}.
It is possible in such a detector to detect and distinguish several
particles traversing the detector simultaneously in different directions.
Of course the Kamiokande detector is much too large to be used as
a single detector element in a colliding beam configuration. The experimental
challenge is to design Cherenkov detectors that are sensitive to the
direction of a traversing particle while at the same time small enough
to be used in in the tracking chamber of a colliding beam experiment. 

As it turns out, a Cherenkov detector is also well suited to address
the second issue of only being sensitive to particles with sufficiently
high energy. The size of the Cherenkov radiation light cone is a measure
of the particle energy: low energy particles produce a large radius
cone while high energy particles produce a narrow light cone. Hence,
a detector that can only detect light cones below a certain radius
is also a detector insensitive to low energy particles.

\section{Conclusions}

In order to study the origin of neutrino masses it is necessary to
explore neutrino interactions at much higher energies than observed
so far. In this paper, we have investigated several colliding beam
scenarios involving neutrinos, colliding them both with other fermions
and with other neutrinos. It is clear that existing technology is
not sufficient to produce neutrino beams with high-enough intensity
to be used in colliding beam configurations. Increases of at least
two orders of magnitude are necessary in the intensity of primary
particles (protons) used to generate neutrino beams. 

If such increases can be achieved, then neutrino colliders become
an attractive possibility. They allow for a first direct measurement
of the neutrino annihilation cross section to $Z$~bosons and several
other electroweak measurements. A neutrino-neutrino collider based
on the LHC will be the first machine sensitive to new physics in the
neutrino sector and possibly the origin of neutrino mass. The high-intensity,
high-energy neutrino beam produced at a muon collider improves upon
the reach of the LHC by several orders of magnitude. Neutrino-neutrino
collisions can be observed already with muon colliders currently under
consideration. And while these neutrino-neutrino collisions might
not be the main motivation for building a muon collider, they nevertheless
provide excellent opportunities to understand neutrinos better and
can easily be incorporated into muon collider designs. 

\bibliographystyle{apsrev}
\clearpage\addcontentsline{toc}{chapter}{\bibname}\bibliography{reference}

\begin{thebibliography}{70}
\expandafter\ifx\csname natexlab\endcsname\relax\def\natexlab#1{#1}\fi
\expandafter\ifx\csname bibnamefont\endcsname\relax
  \def\bibnamefont#1{#1}\fi
\expandafter\ifx\csname bibfnamefont\endcsname\relax
  \def\bibfnamefont#1{#1}\fi
\expandafter\ifx\csname citenamefont\endcsname\relax
  \def\citenamefont#1{#1}\fi
\expandafter\ifx\csname url\endcsname\relax
  \def\url#1{\texttt{#1}}\fi
\expandafter\ifx\csname urlprefix\endcsname\relax\def\urlprefix{URL }\fi
\providecommand{\bibinfo}[2]{#2}
\providecommand{\eprint}[2][]{\url{#2}}

\bibitem[{\citenamefont{Danby et~al.}(1962)}]{Danby:1962nd}
\bibinfo{author}{\bibfnamefont{G.}~\bibnamefont{Danby}} \bibnamefont{et~al.},
  \bibinfo{journal}{Phys. Rev. Lett.} \textbf{\bibinfo{volume}{9}},
  \bibinfo{pages}{36} (\bibinfo{year}{1962}).

\bibitem[{\citenamefont{Fukuda et~al.}(1998)}]{Fukuda:1998mi}
\bibinfo{author}{\bibfnamefont{Y.}~\bibnamefont{Fukuda}} \bibnamefont{et~al.}
  (\bibinfo{collaboration}{D0}), \bibinfo{journal}{Phys. Rev. Lett.}
  \textbf{\bibinfo{volume}{81}}, \bibinfo{pages}{1562} (\bibinfo{year}{1998}),
  \eprint{hep-ex/0406031}.

\bibitem[{\citenamefont{Davis et~al.}(1968)\citenamefont{Davis, Harmer, and
  Hoffman}}]{Davis:1968cp}
\bibinfo{author}{\bibfnamefont{J.}~\bibnamefont{Davis},
  \bibfnamefont{Raymond}}, \bibinfo{author}{\bibfnamefont{D.~S.}
  \bibnamefont{Harmer}}, \bibnamefont{and}
  \bibinfo{author}{\bibfnamefont{K.~C.} \bibnamefont{Hoffman}},
  \bibinfo{journal}{Phys. Rev. Lett.} \textbf{\bibinfo{volume}{20}},
  \bibinfo{pages}{1205} (\bibinfo{year}{1968}).

\bibitem[{\citenamefont{Ahmad et~al.}(2002)}]{SNO:2002}
\bibinfo{author}{\bibfnamefont{Q.}~\bibnamefont{Ahmad}} \bibnamefont{et~al.},
  \bibinfo{journal}{Phys.Rev.Lett.} \textbf{\bibinfo{volume}{89}},
  \bibinfo{pages}{011301} (\bibinfo{year}{2002}).

\bibitem[{\citenamefont{Aliu et~al.}(2005)}]{Aliu:2004sq}
\bibinfo{author}{\bibfnamefont{E.}~\bibnamefont{Aliu}} \bibnamefont{et~al.}
  (\bibinfo{collaboration}{K2K}), \bibinfo{journal}{Phys. Rev. Lett.}
  \textbf{\bibinfo{volume}{94}}, \bibinfo{pages}{081802}
  (\bibinfo{year}{2005}), \eprint{hep-ex/0411038}.

\bibitem[{\citenamefont{Michael et~al.}(2006)}]{Michael:2006rx}
\bibinfo{author}{\bibfnamefont{D.~G.} \bibnamefont{Michael}}
  \bibnamefont{et~al.} (\bibinfo{collaboration}{MINOS}),
  \bibinfo{journal}{Phys. Rev. Lett.} \textbf{\bibinfo{volume}{97}},
  \bibinfo{pages}{191801} (\bibinfo{year}{2006}), \eprint{hep-ex/0607088}.

\bibitem[{\citenamefont{Higgs}(1964)}]{Higgs:1964ia}
\bibinfo{author}{\bibfnamefont{P.~W.} \bibnamefont{Higgs}},
  \bibinfo{journal}{Phys. Lett.} \textbf{\bibinfo{volume}{12}},
  \bibinfo{pages}{132} (\bibinfo{year}{1964}).

\bibitem[{\citenamefont{Abazov et~al.}(2005{\natexlab{a}})}]{Abazov:2004jy}
\bibinfo{author}{\bibfnamefont{V.~M.} \bibnamefont{Abazov}}
  \bibnamefont{et~al.} (\bibinfo{collaboration}{D0}), \bibinfo{journal}{Phys.
  Rev. Lett.} \textbf{\bibinfo{volume}{94}}, \bibinfo{pages}{091802}
  (\bibinfo{year}{2005}{\natexlab{a}}), \eprint{hep-ex/0410062}.

\bibitem[{\citenamefont{Abazov et~al.}(2006{\natexlab{a}})}]{Abazov:2005un}
\bibinfo{author}{\bibfnamefont{V.~M.} \bibnamefont{Abazov}}
  \bibnamefont{et~al.} (\bibinfo{collaboration}{D0}), \bibinfo{journal}{Phys.
  Rev. Lett.} \textbf{\bibinfo{volume}{96}}, \bibinfo{pages}{011801}
  (\bibinfo{year}{2006}{\natexlab{a}}), \eprint{hep-ex/0508054}.

\bibitem[{\citenamefont{Abazov et~al.}(2006{\natexlab{b}})}]{Abazov:2006hn}
\bibinfo{author}{\bibfnamefont{V.~M.} \bibnamefont{Abazov}}
  \bibnamefont{et~al.} (\bibinfo{collaboration}{D0}), \bibinfo{journal}{Phys.
  Rev. Lett.} \textbf{\bibinfo{volume}{97}}, \bibinfo{pages}{151804}
  (\bibinfo{year}{2006}{\natexlab{b}}), \eprint{hep-ex/0607032}.

\bibitem[{\citenamefont{Abazov et~al.}(2006{\natexlab{c}})}]{Abazov:2006pt}
\bibinfo{author}{\bibfnamefont{V.~M.} \bibnamefont{Abazov}}
  \bibnamefont{et~al.} (\bibinfo{collaboration}{D0}), \bibinfo{journal}{Phys.
  Rev. Lett.} \textbf{\bibinfo{volume}{97}}, \bibinfo{pages}{161803}
  (\bibinfo{year}{2006}{\natexlab{c}}), \eprint{hep-ex/0607022}.

\bibitem[{\citenamefont{Abulencia
  et~al.}(2006{\natexlab{a}})}]{Abulencia:2006aj}
\bibinfo{author}{\bibfnamefont{A.}~\bibnamefont{Abulencia}}
  \bibnamefont{et~al.} (\bibinfo{collaboration}{CDF}), \bibinfo{journal}{Phys.
  Rev. Lett.} \textbf{\bibinfo{volume}{97}}, \bibinfo{pages}{081802}
  (\bibinfo{year}{2006}{\natexlab{a}}), \eprint{hep-ex/0605124}.

\bibitem[{\citenamefont{Acosta et~al.}(2005{\natexlab{a}})}]{Acosta:2005ga}
\bibinfo{author}{\bibfnamefont{D.}~\bibnamefont{Acosta}} \bibnamefont{et~al.}
  (\bibinfo{collaboration}{CDF}), \bibinfo{journal}{Phys. Rev. Lett.}
  \textbf{\bibinfo{volume}{95}}, \bibinfo{pages}{051801}
  (\bibinfo{year}{2005}{\natexlab{a}}), \eprint{hep-ex/0503039}.

\bibitem[{\citenamefont{Abazov et~al.}(2004)}]{Abazov:2004au}
\bibinfo{author}{\bibfnamefont{V.~M.} \bibnamefont{Abazov}}
  \bibnamefont{et~al.} (\bibinfo{collaboration}{D0}), \bibinfo{journal}{Phys.
  Rev. Lett.} \textbf{\bibinfo{volume}{93}}, \bibinfo{pages}{141801}
  (\bibinfo{year}{2004}), \eprint{hep-ex/0404015}.

\bibitem[{\citenamefont{Abazov et~al.}(2005{\natexlab{b}})}]{Abazov:2005yr}
\bibinfo{author}{\bibfnamefont{V.~M.} \bibnamefont{Abazov}}
  \bibnamefont{et~al.} (\bibinfo{collaboration}{D0}), \bibinfo{journal}{Phys.
  Rev. Lett.} \textbf{\bibinfo{volume}{95}}, \bibinfo{pages}{151801}
  (\bibinfo{year}{2005}{\natexlab{b}}), \eprint{hep-ex/0504018}.

\bibitem[{\citenamefont{Abazov et~al.}(2006{\natexlab{d}})}]{Abazov:2006ih}
\bibinfo{author}{\bibfnamefont{V.~M.} \bibnamefont{Abazov}}
  \bibnamefont{et~al.} (\bibinfo{collaboration}{D0}), \bibinfo{journal}{Phys.
  Rev. Lett.} \textbf{\bibinfo{volume}{97}}, \bibinfo{pages}{121802}
  (\bibinfo{year}{2006}{\natexlab{d}}), \eprint{hep-ex/0605009}.

\bibitem[{\citenamefont{Abulencia
  et~al.}(2006{\natexlab{b}})}]{Abulencia:2005jd}
\bibinfo{author}{\bibfnamefont{A.}~\bibnamefont{Abulencia}}
  \bibnamefont{et~al.} (\bibinfo{collaboration}{CDF}), \bibinfo{journal}{Phys.
  Rev. Lett.} \textbf{\bibinfo{volume}{96}}, \bibinfo{pages}{042003}
  (\bibinfo{year}{2006}{\natexlab{b}}), \eprint{hep-ex/0510065}.

\bibitem[{\citenamefont{Abulencia
  et~al.}(2006{\natexlab{c}})}]{Abulencia:2005kq}
\bibinfo{author}{\bibfnamefont{A.}~\bibnamefont{Abulencia}}
  \bibnamefont{et~al.} (\bibinfo{collaboration}{CDF}), \bibinfo{journal}{Phys.
  Rev. Lett.} \textbf{\bibinfo{volume}{96}}, \bibinfo{pages}{011802}
  (\bibinfo{year}{2006}{\natexlab{c}}), \eprint{hep-ex/0508051}.

\bibitem[{\citenamefont{Acosta et~al.}(2004)}]{Acosta:2004uj}
\bibinfo{author}{\bibfnamefont{D.}~\bibnamefont{Acosta}} \bibnamefont{et~al.}
  (\bibinfo{collaboration}{CDF}), \bibinfo{journal}{Phys. Rev. Lett.}
  \textbf{\bibinfo{volume}{93}}, \bibinfo{pages}{221802}
  (\bibinfo{year}{2004}), \eprint{hep-ex/0406073}.

\bibitem[{\citenamefont{Acosta et~al.}(2005{\natexlab{b}})}]{Acosta:2005bk}
\bibinfo{author}{\bibfnamefont{D.}~\bibnamefont{Acosta}} \bibnamefont{et~al.}
  (\bibinfo{collaboration}{CDF}), \bibinfo{journal}{Phys. Rev.}
  \textbf{\bibinfo{volume}{D72}}, \bibinfo{pages}{072004}
  (\bibinfo{year}{2005}{\natexlab{b}}), \eprint{hep-ex/0506042}.

\bibitem[{\citenamefont{Acosta et~al.}(2005{\natexlab{c}})}]{Acosta:2005np}
\bibinfo{author}{\bibfnamefont{D.}~\bibnamefont{Acosta}} \bibnamefont{et~al.}
  (\bibinfo{collaboration}{CDF}), \bibinfo{journal}{Phys. Rev. Lett.}
  \textbf{\bibinfo{volume}{95}}, \bibinfo{pages}{071801}
  (\bibinfo{year}{2005}{\natexlab{c}}), \eprint{hep-ex/0503004}.

\bibitem[{\citenamefont{CMS}(1994)}]{unknown:1994pu}
\bibinfo{author}{\bibnamefont{CMS}} (\bibinfo{year}{1994}),
  \bibinfo{note}{cERN-LHCC-94-38}.

\bibitem[{\citenamefont{ATLAS}(1999{\natexlab{a}})}]{unknown:1999fq}
\bibinfo{author}{\bibnamefont{ATLAS}} (\bibinfo{year}{1999}{\natexlab{a}}),
  \bibinfo{note}{cERN-LHCC-99-14}.

\bibitem[{\citenamefont{ATLAS}(1999{\natexlab{b}})}]{unknown:1999fr}
\bibinfo{author}{\bibnamefont{ATLAS}} (\bibinfo{year}{1999}{\natexlab{b}}),
  \bibinfo{note}{cERN-LHCC-99-15}.

\bibitem[{\citenamefont{Armstrong et~al.}(1994)}]{Armstrong:1994it}
\bibinfo{author}{\bibfnamefont{W.~W.} \bibnamefont{Armstrong}}
  \bibnamefont{et~al.} (\bibinfo{collaboration}{ATLAS}) (\bibinfo{year}{1994}),
  \bibinfo{note}{cERN-LHCC-94-43}.

\bibitem[{\citenamefont{Mohapatra and Smirnov}(2006)}]{Mohapatra:2006gs}
\bibinfo{author}{\bibfnamefont{R.~N.} \bibnamefont{Mohapatra}}
  \bibnamefont{and} \bibinfo{author}{\bibfnamefont{A.~Y.}
  \bibnamefont{Smirnov}}, \bibinfo{journal}{Ann. Rev. Nucl. Part. Sci.}
  \textbf{\bibinfo{volume}{56}}, \bibinfo{pages}{569} (\bibinfo{year}{2006}),
  \eprint{hep-ph/0603118}.

\bibitem[{\citenamefont{Caldwell and Mohapatra}(1993)}]{Caldwell:1993kn}
\bibinfo{author}{\bibfnamefont{D.~O.} \bibnamefont{Caldwell}} \bibnamefont{and}
  \bibinfo{author}{\bibfnamefont{R.~N.} \bibnamefont{Mohapatra}},
  \bibinfo{journal}{Phys. Rev.} \textbf{\bibinfo{volume}{D48}},
  \bibinfo{pages}{3259} (\bibinfo{year}{1993}).

\bibitem[{\citenamefont{Ma and Roy}(1995)}]{Ma:1995gf}
\bibinfo{author}{\bibfnamefont{E.}~\bibnamefont{Ma}} \bibnamefont{and}
  \bibinfo{author}{\bibfnamefont{P.}~\bibnamefont{Roy}},
  \bibinfo{journal}{Phys. Rev.} \textbf{\bibinfo{volume}{D52}},
  \bibinfo{pages}{R4780} (\bibinfo{year}{1995}), \eprint{hep-ph/9504342}.

\bibitem[{\citenamefont{Berezhiani and Mohapatra}(1995)}]{Berezhiani:1995yi}
\bibinfo{author}{\bibfnamefont{Z.~G.} \bibnamefont{Berezhiani}}
  \bibnamefont{and} \bibinfo{author}{\bibfnamefont{R.~N.}
  \bibnamefont{Mohapatra}}, \bibinfo{journal}{Phys. Rev.}
  \textbf{\bibinfo{volume}{D52}}, \bibinfo{pages}{6607} (\bibinfo{year}{1995}),
  \eprint{hep-ph/9505385}.

\bibitem[{\citenamefont{Benakli and Smirnov}(1997)}]{Benakli:1997iu}
\bibinfo{author}{\bibfnamefont{K.}~\bibnamefont{Benakli}} \bibnamefont{and}
  \bibinfo{author}{\bibfnamefont{A.~Y.} \bibnamefont{Smirnov}},
  \bibinfo{journal}{Phys. Rev. Lett.} \textbf{\bibinfo{volume}{79}},
  \bibinfo{pages}{4314} (\bibinfo{year}{1997}), \eprint{hep-ph/9703465}.

\bibitem[{\citenamefont{Eidelman et~al.}(2004)\citenamefont{Eidelman, {Hayes},
  {Olive}, {Aguilar-Benitez}, {Amsler}, {Asner}, {Babu}, {Barnett}, {Beringer},
  {Burchat} et~al.}}]{PDBook}
\bibinfo{author}{\bibfnamefont{S.}~\bibnamefont{Eidelman}},
  \bibinfo{author}{\bibfnamefont{K.}~\bibnamefont{{Hayes}}},
  \bibinfo{author}{\bibfnamefont{K.}~\bibnamefont{{Olive}}},
  \bibinfo{author}{\bibfnamefont{M.}~\bibnamefont{{Aguilar-Benitez}}},
  \bibinfo{author}{\bibfnamefont{C.}~\bibnamefont{{Amsler}}},
  \bibinfo{author}{\bibfnamefont{D.}~\bibnamefont{{Asner}}},
  \bibinfo{author}{\bibfnamefont{K.}~\bibnamefont{{Babu}}},
  \bibinfo{author}{\bibfnamefont{R.}~\bibnamefont{{Barnett}}},
  \bibinfo{author}{\bibfnamefont{J.}~\bibnamefont{{Beringer}}},
  \bibinfo{author}{\bibfnamefont{P.}~\bibnamefont{{Burchat}}},
  \bibnamefont{et~al.}, \bibinfo{journal}{Physics Letters B}
  \textbf{\bibinfo{volume}{592}} (\bibinfo{year}{2004}).

\bibitem[{\citenamefont{Barger et~al.}(2003)\citenamefont{Barger, Marfatia, and
  Whisnant}}]{Barger:2003qi}
\bibinfo{author}{\bibfnamefont{V.}~\bibnamefont{Barger}},
  \bibinfo{author}{\bibfnamefont{D.}~\bibnamefont{Marfatia}}, \bibnamefont{and}
  \bibinfo{author}{\bibfnamefont{K.}~\bibnamefont{Whisnant}},
  \bibinfo{journal}{Int. J. Mod. Phys.} \textbf{\bibinfo{volume}{E12}},
  \bibinfo{pages}{569} (\bibinfo{year}{2003}), \eprint{hep-ph/0308123}.

\bibitem[{\citenamefont{Ables et~al.}(1995)}]{Ables:1995wq}
\bibinfo{author}{\bibfnamefont{E.}~\bibnamefont{Ables}} \bibnamefont{et~al.}
  (\bibinfo{collaboration}{MINOS}) (\bibinfo{year}{1995}),
  \bibinfo{note}{fERMILAB-PROPOSAL-0875}.

\bibitem[{\citenamefont{Acquistapace et~al.}(1998)}]{Acquistapace:1998rv}
\bibinfo{author}{\bibfnamefont{G.}~\bibnamefont{Acquistapace}}
  \bibnamefont{et~al.} (\bibinfo{year}{1998}), \bibinfo{note}{cERN-98-02}.

\bibitem[{\citenamefont{Han and Zhang}(2006)}]{Han:2006ip}
\bibinfo{author}{\bibfnamefont{T.}~\bibnamefont{Han}} \bibnamefont{and}
  \bibinfo{author}{\bibfnamefont{B.}~\bibnamefont{Zhang}},
  \bibinfo{journal}{Phys. Rev. Lett.} \textbf{\bibinfo{volume}{97}},
  \bibinfo{pages}{171804} (\bibinfo{year}{2006}), \eprint{hep-ph/0604064}.

\bibitem[{\citenamefont{Avignone et~al.}(2007)\citenamefont{Avignone, Elliott,
  and Engel}}]{Avignone:2007fu}
\bibinfo{author}{\bibfnamefont{I.}~\bibnamefont{Avignone},
  \bibfnamefont{Frank~T.}}, \bibinfo{author}{\bibfnamefont{S.~R.}
  \bibnamefont{Elliott}}, \bibnamefont{and}
  \bibinfo{author}{\bibfnamefont{J.}~\bibnamefont{Engel}}
  (\bibinfo{year}{2007}), \eprint{arXiv:0708.1033 [nucl-ex]}.

\bibitem[{\citenamefont{Schwartz}(1960)}]{Schwartz:1960hg}
\bibinfo{author}{\bibfnamefont{M.}~\bibnamefont{Schwartz}},
  \bibinfo{journal}{Phys. Rev. Lett.} \textbf{\bibinfo{volume}{4}},
  \bibinfo{pages}{306} (\bibinfo{year}{1960}).

\bibitem[{\citenamefont{Pontecorvo}(1960)}]{Pontecorvo:1959sn}
\bibinfo{author}{\bibfnamefont{B.}~\bibnamefont{Pontecorvo}},
  \bibinfo{journal}{Sov. Phys. JETP} \textbf{\bibinfo{volume}{10}},
  \bibinfo{pages}{1236} (\bibinfo{year}{1960}).

\bibitem[{\citenamefont{Onengut et~al.}(2006)}]{Onengut:2005kv}
\bibinfo{author}{\bibfnamefont{G.}~\bibnamefont{Onengut}} \bibnamefont{et~al.}
  (\bibinfo{collaboration}{CHORUS}), \bibinfo{journal}{Phys. Lett.}
  \textbf{\bibinfo{volume}{B632}}, \bibinfo{pages}{65} (\bibinfo{year}{2006}).

\bibitem[{\citenamefont{Seligman et~al.}(1997)}]{Seligman:1997mc}
\bibinfo{author}{\bibfnamefont{W.~G.} \bibnamefont{Seligman}}
  \bibnamefont{et~al.}, \bibinfo{journal}{Phys. Rev. Lett.}
  \textbf{\bibinfo{volume}{79}}, \bibinfo{pages}{1213} (\bibinfo{year}{1997}).

\bibitem[{\citenamefont{Aivazis et~al.}(1994)\citenamefont{Aivazis, Collins,
  Olness, and Tung}}]{Aivazis:1993pi}
\bibinfo{author}{\bibfnamefont{M.~A.~G.} \bibnamefont{Aivazis}},
  \bibinfo{author}{\bibfnamefont{J.~C.} \bibnamefont{Collins}},
  \bibinfo{author}{\bibfnamefont{F.~I.} \bibnamefont{Olness}},
  \bibnamefont{and} \bibinfo{author}{\bibfnamefont{W.-K.} \bibnamefont{Tung}},
  \bibinfo{journal}{Phys. Rev.} \textbf{\bibinfo{volume}{D50}},
  \bibinfo{pages}{3102} (\bibinfo{year}{1994}), \eprint{hep-ph/9312319}.

\bibitem[{\citenamefont{Rabinowitz et~al.}(1993)}]{Rabinowitz:1993xx}
\bibinfo{author}{\bibfnamefont{S.~A.} \bibnamefont{Rabinowitz}}
  \bibnamefont{et~al.}, \bibinfo{journal}{Phys. Rev. Lett.}
  \textbf{\bibinfo{volume}{70}}, \bibinfo{pages}{134} (\bibinfo{year}{1993}).

\bibitem[{\citenamefont{Abramowicz et~al.}(1985)}]{Abramowicz:1985xg}
\bibinfo{author}{\bibfnamefont{H.}~\bibnamefont{Abramowicz}}
  \bibnamefont{et~al.}, \bibinfo{journal}{Z. Phys.}
  \textbf{\bibinfo{volume}{C28}}, \bibinfo{pages}{51} (\bibinfo{year}{1985}).

\bibitem[{\citenamefont{Jones et~al.}(1989)}]{Jones:1987gk}
\bibinfo{author}{\bibfnamefont{G.~T.} \bibnamefont{Jones}}
  \bibnamefont{et~al.}, \bibinfo{journal}{Z. Phys.}
  \textbf{\bibinfo{volume}{C44}}, \bibinfo{pages}{379} (\bibinfo{year}{1989}).

\bibitem[{\citenamefont{Vogel and Engel}(1989)}]{Vogel:1989iv}
\bibinfo{author}{\bibfnamefont{P.}~\bibnamefont{Vogel}} \bibnamefont{and}
  \bibinfo{author}{\bibfnamefont{J.}~\bibnamefont{Engel}},
  \bibinfo{journal}{Phys. Rev.} \textbf{\bibinfo{volume}{D39}},
  \bibinfo{pages}{3378} (\bibinfo{year}{1989}).

\bibitem[{\citenamefont{Abramowicz and Caldwell}(1999)}]{Abramowicz:1998ii}
\bibinfo{author}{\bibfnamefont{H.}~\bibnamefont{Abramowicz}} \bibnamefont{and}
  \bibinfo{author}{\bibfnamefont{A.}~\bibnamefont{Caldwell}},
  \bibinfo{journal}{Rev. Mod. Phys.} \textbf{\bibinfo{volume}{71}},
  \bibinfo{pages}{1275} (\bibinfo{year}{1999}), \eprint{hep-ex/9903037}.

\bibitem[{\citenamefont{Pati and Salam}(1974)}]{Pati:1974yy}
\bibinfo{author}{\bibfnamefont{J.~C.} \bibnamefont{Pati}} \bibnamefont{and}
  \bibinfo{author}{\bibfnamefont{A.}~\bibnamefont{Salam}},
  \bibinfo{journal}{Phys. Rev.} \textbf{\bibinfo{volume}{D10}},
  \bibinfo{pages}{275} (\bibinfo{year}{1974}).

\bibitem[{\citenamefont{Mohapatra and Senjanovic}(1980)}]{Mohapatra:1979ia}
\bibinfo{author}{\bibfnamefont{R.~N.} \bibnamefont{Mohapatra}}
  \bibnamefont{and}
  \bibinfo{author}{\bibfnamefont{G.}~\bibnamefont{Senjanovic}},
  \bibinfo{journal}{Phys. Rev. Lett.} \textbf{\bibinfo{volume}{44}},
  \bibinfo{pages}{912} (\bibinfo{year}{1980}).

\bibitem[{\citenamefont{Mohapatra and Senjanovic}(1981)}]{Mohapatra:1980yp}
\bibinfo{author}{\bibfnamefont{R.~N.} \bibnamefont{Mohapatra}}
  \bibnamefont{and}
  \bibinfo{author}{\bibfnamefont{G.}~\bibnamefont{Senjanovic}},
  \bibinfo{journal}{Phys. Rev.} \textbf{\bibinfo{volume}{D23}},
  \bibinfo{pages}{165} (\bibinfo{year}{1981}).

\bibitem[{\citenamefont{Ma}(2001)}]{PhysRevLett.86.2502}
\bibinfo{author}{\bibfnamefont{E.}~\bibnamefont{Ma}}, \bibinfo{journal}{Phys.
  Rev. Lett.} \textbf{\bibinfo{volume}{86}}, \bibinfo{pages}{2502}
  (\bibinfo{year}{2001}).

\bibitem[{\citenamefont{Heusch}(2000)}]{Heusch:2000cp}
\bibinfo{author}{\bibfnamefont{C.~A.} \bibnamefont{Heusch}},
  \bibinfo{journal}{Batavia 2000, Physics and experiments with future linear e+
  e- colliders} pp. \bibinfo{pages}{627--630} (\bibinfo{year}{2000}),
  \bibinfo{note}{prepared for 5th International Linear Collider Workshop (LCWS
  2000), Fermilab, Batavia, Illinois, 24-28 Oct 2000}.

\bibitem[{\citenamefont{Vuopionpera}(1995)}]{Vuopionpera:1994kr}
\bibinfo{author}{\bibfnamefont{R.}~\bibnamefont{Vuopionpera}},
  \bibinfo{journal}{Z. Phys.} \textbf{\bibinfo{volume}{C65}},
  \bibinfo{pages}{311} (\bibinfo{year}{1995}).

\bibitem[{\citenamefont{Abreu et~al.}(1997)}]{Abreu:1996vd}
\bibinfo{author}{\bibfnamefont{P.}~\bibnamefont{Abreu}} \bibnamefont{et~al.}
  (\bibinfo{collaboration}{DELPHI}), \bibinfo{journal}{Z. Phys.}
  \textbf{\bibinfo{volume}{C74}}, \bibinfo{pages}{577} (\bibinfo{year}{1997}).

\bibitem[{\citenamefont{Achard et~al.}(2004)}]{Achard:2003tx}
\bibinfo{author}{\bibfnamefont{P.}~\bibnamefont{Achard}} \bibnamefont{et~al.}
  (\bibinfo{collaboration}{L3}), \bibinfo{journal}{Phys. Lett.}
  \textbf{\bibinfo{volume}{B587}}, \bibinfo{pages}{16} (\bibinfo{year}{2004}),
  \eprint{hep-ex/0402002}.

\bibitem[{\citenamefont{{The ALEPH, DELPHI, L3, OPAL, SLD Collaborations, the
  LEP Electroweak Working Group, the SLD Electroweak and Heavy Flavour
  Groups}}(2006)}]{Z-Pole}
\bibinfo{author}{\bibnamefont{{The ALEPH, DELPHI, L3, OPAL, SLD Collaborations,
  the LEP Electroweak Working Group, the SLD Electroweak and Heavy Flavour
  Groups}}}, \bibinfo{journal}{Phys. Rept.} \textbf{\bibinfo{volume}{427}},
  \bibinfo{pages}{257} (\bibinfo{year}{2006}), \eprint{hep-ex/0509008}.

\bibitem[{\citenamefont{McFarland et~al.}(1995)}]{McFarland:1995sr}
\bibinfo{author}{\bibfnamefont{K.~S.} \bibnamefont{McFarland}}
  \bibnamefont{et~al.}, \bibinfo{journal}{Phys. Rev. Lett.}
  \textbf{\bibinfo{volume}{75}}, \bibinfo{pages}{3993} (\bibinfo{year}{1995}),
  \eprint{hep-ex/9506007}.

\bibitem[{\citenamefont{Noumi et~al.}(1997)}]{Noumi:1997}
\bibinfo{author}{\bibfnamefont{H.}~\bibnamefont{Noumi}} \bibnamefont{et~al.},
  \bibinfo{journal}{Nucl. Instr. and Meth.} \textbf{\bibinfo{volume}{A398}}
  (\bibinfo{year}{1997}).

\bibitem[{\citenamefont{Crane et~al.}(1995)}]{Crane:1995ky}
\bibinfo{author}{\bibfnamefont{D.~A.} \bibnamefont{Crane}} \bibnamefont{et~al.}
  (\bibinfo{collaboration}{NuMI Beam Group}) (\bibinfo{year}{1995}),
  \bibinfo{note}{fERMILAB-TM-1946}.

\bibitem[{\citenamefont{Koba et~al.}(2003)}]{Koba:2003sj}
\bibinfo{author}{\bibfnamefont{K.}~\bibnamefont{Koba}} \bibnamefont{et~al.}
  (\bibinfo{year}{2003}), \bibinfo{note}{presented at Particle Accelerator
  Conference (PAC 03), Portland, Oregon, 12-16 May 2003}.

\bibitem[{\citenamefont{Bonesini et~al.}(2001)\citenamefont{Bonesini,
  Marchionni, Pietropaolo, and Tabarelli~de Fatis}}]{Bonesini:2001iz}
\bibinfo{author}{\bibfnamefont{M.}~\bibnamefont{Bonesini}},
  \bibinfo{author}{\bibfnamefont{A.}~\bibnamefont{Marchionni}},
  \bibinfo{author}{\bibfnamefont{F.}~\bibnamefont{Pietropaolo}},
  \bibnamefont{and}
  \bibinfo{author}{\bibfnamefont{T.}~\bibnamefont{Tabarelli~de Fatis}},
  \bibinfo{journal}{Eur. Phys. J.} \textbf{\bibinfo{volume}{C20}},
  \bibinfo{pages}{13} (\bibinfo{year}{2001}), \eprint{hep-ph/0101163}.

\bibitem[{\citenamefont{Geer}(1998)}]{Geer:1997iz}
\bibinfo{author}{\bibfnamefont{S.}~\bibnamefont{Geer}}, \bibinfo{journal}{Phys.
  Rev.} \textbf{\bibinfo{volume}{D57}}, \bibinfo{pages}{6989}
  (\bibinfo{year}{1998}), \eprint{hep-ph/9712290}.

\bibitem[{\citenamefont{Group}()}]{nufact}
\bibinfo{author}{\bibfnamefont{E.~N.} \bibnamefont{Group}},
  \emph{\bibinfo{title}{Neutrino factory study}},
  \urlprefix\url{http://nfwg.home.cern.ch/nfwg/nufactwg/nufactwg.html}.

\bibitem[{\citenamefont{Albright et~al.}(2000)}]{Albright:2000xi}
\bibinfo{author}{\bibfnamefont{C.}~\bibnamefont{Albright}} \bibnamefont{et~al.}
  (\bibinfo{year}{2000}), \eprint{hep-ex/0008064}.

\bibitem[{\citenamefont{Albright et~al.}(2004)}]{Albright:2004iw}
\bibinfo{author}{\bibfnamefont{C.}~\bibnamefont{Albright}} \bibnamefont{et~al.}
  (\bibinfo{collaboration}{Neutrino Factory/Muon Collider})
  (\bibinfo{year}{2004}), \eprint{physics/0411123}.

\bibitem[{\citenamefont{Maltoni and Stelzer}(2003)}]{Maltoni:2002qb}
\bibinfo{author}{\bibfnamefont{F.}~\bibnamefont{Maltoni}} \bibnamefont{and}
  \bibinfo{author}{\bibfnamefont{T.}~\bibnamefont{Stelzer}},
  \bibinfo{journal}{JHEP} \textbf{\bibinfo{volume}{02}}, \bibinfo{pages}{027}
  (\bibinfo{year}{2003}), \eprint{hep-ph/0208156}.

\bibitem[{\citenamefont{Ahmed et~al.}(1995)}]{Ahmed:1995fd}
\bibinfo{author}{\bibfnamefont{T.}~\bibnamefont{Ahmed}} \bibnamefont{et~al.}
  (\bibinfo{collaboration}{H1}), \bibinfo{journal}{Nucl. Phys.}
  \textbf{\bibinfo{volume}{B439}}, \bibinfo{pages}{471} (\bibinfo{year}{1995}),
  \eprint{hep-ex/9503001}.

\bibitem[{\citenamefont{Derrick et~al.}(1995)}]{Derrick:1994sz}
\bibinfo{author}{\bibfnamefont{M.}~\bibnamefont{Derrick}} \bibnamefont{et~al.}
  (\bibinfo{collaboration}{ZEUS}), \bibinfo{journal}{Z. Phys.}
  \textbf{\bibinfo{volume}{C65}}, \bibinfo{pages}{379} (\bibinfo{year}{1995}).

\bibitem[{\citenamefont{Finley et~al.}(1999)\citenamefont{Finley, Geer, and
  Sims}}]{Finley:1999yt}
\bibinfo{author}{\bibfnamefont{D.}~\bibnamefont{Finley}},
  \bibinfo{author}{\bibfnamefont{S.}~\bibnamefont{Geer}}, \bibnamefont{and}
  \bibinfo{author}{\bibfnamefont{J.}~\bibnamefont{Sims}}
  (\bibinfo{year}{1999}), \bibinfo{note}{fERMILAB-TM-2072}.

\bibitem[{\citenamefont{Ahn et~al.}(1999)}]{Ahn:1999kd}
\bibinfo{author}{\bibfnamefont{S.-C.} \bibnamefont{Ahn}} \bibnamefont{et~al.}
  (\bibinfo{year}{1999}), \bibinfo{note}{fERMILAB-FN-0677}.

\bibitem[{\citenamefont{Kodama et~al.}(2001)}]{donutDiscovery}
\bibinfo{author}{\bibfnamefont{K.}~\bibnamefont{Kodama}} \bibnamefont{et~al.}
  (\bibinfo{collaboration}{DONUT}), \bibinfo{journal}{Phys. Lett.}
  \textbf{\bibinfo{volume}{B504}}, \bibinfo{pages}{218} (\bibinfo{year}{2001}),
  \eprint{hep-ex/0012035}.

\end{thebibliography}

\end{document}